\documentclass[twocolumn,english,aps,pre,floats,superscriptaddress]{revtex4}

\usepackage[pdftex]{graphicx}
\usepackage{latexsym,amsmath}
\usepackage{amsbsy}
\usepackage[usenames]{color}
\usepackage{amssymb}
\usepackage{graphicx}

\begin{document}
\title{Uncovering the hidden geometry behind metabolic networks}

\author{M. \'Angeles Serrano}
\affiliation{Departament de Qu\'{\i}mica F\'{\i}sica, Universitat de Barcelona, Barcelona, Spain}

\author{Mari\'an Bogu\~{n}\'a}
\affiliation{Departament de F{\'\i}sica Fonamental, Universitat de Barcelona, Barcelona, Spain}

\author{Francesc Sagu\'es}
\affiliation{Departament de Qu\'{\i}mica F\'{\i}sica, Universitat de Barcelona, Barcelona, Spain}

\date{\today}

\begin{abstract}Metabolism is a fascinating cell machinery underlying life and disease and genome-scale reconstructions provide us with a captivating view of its complexity. However, deciphering the relationship between metabolic structure and function remains a major challenge. In particular, turning observed structural regularities into organizing principles underlying systemic functions is a crucial task that can be significantly addressed after endowing complex network representations of metabolism with the notion of geometric distance. Here, we design a cartographic map of metabolic networks by embedding them into a simple geometry that provides a natural explanation for their observed network topology and that codifies node proximity as a measure of hidden structural similarities. We assume a simple and general connectivity law that gives more probability of interaction to metabolite/reaction pairs which are closer in the hidden space. Remarkably, we find an astonishing congruency between the architecture of E. coli and human cell metabolisms and the underlying geometry. In addition, the formalism unveils a backbone-like structure of connected biochemical pathways on the basis of a quantitative cross-talk. Pathways thus acquire a new perspective which challenges their classical view as self-contained functional units. 
\end{abstract}

\maketitle

Cells are self-organized entities that carry-out specialized tasks at different interrelated omic-levels~\cite{Palsson:2010} involving different actors, from codifying genes to energy-carrier or constitutive metabolites. A key towards understanding this complex architecture at a systems level is provided by reliable genome-wide reconstructions of the set of biochemical reactions that underly the functional cell machinery~\cite{Palsson:2006}. Such reconstructions can be analyzed using tools and techniques from complex networks theory~\cite{Dorogovtsev:2002,Albert:2002,Newman:2003}, a discipline that is being used in the characterization of biological, chemical, infrastructural, technological or social-based systems of complex relationships~\cite{Dorogovtsev:2003,newmanbook}. More precisely, nodes in metabolic networks account for either metabolites or reactions, while links represent the interactions among them. Apart from providing a large-scale organizational picture, these network-based representations have permitted to analyze sensible issues in cellular metabolism, like flux balances~\cite{Edwards:2001uq,Almaas:2004}, regulation~\cite{Stelling:2002}, robustness~\cite{Smart:2008}, or reaction reliability~\cite{Serrano:2011b}.

The advantage of using network-based representations, in whatever context we employ them, may be arguably questioned by the fact that complex networks are customarily modeled as pure topological constructions lacking a true geometric measure of separation among nodes. This is aggravated by the fact that complex networks have the small-world property~\cite{Watts:1998}, meaning that every pair of nodes in the system are very close in topological distance. This is an important and obvious degeneracy if we think in terms of optimizing routing or transportation strategies in man-engineered networks, but can be equally crucial when referring to the description of the metabolic functioning at a single cell level. As a matter of fact, the related attempt of separating nodes into communities, that has been already pursued in different contexts~\cite{Fortunato201075} and, in particular, applied to metabolic networks~\cite{Guimera:2005b}, has proven to be an extremely difficult task. Classical community detection approaches turn out to be {\it a posteriori} classification methods, and do not provide insights into any potential connectivity law underlying the observed topology. These questions could be significantly addressed by quantifying the abstract concept of node proximity in terms of a metric distance which could be combined into a simple and general probabilistic connectivity law. Such a biochemical connectivity law, relying on metric distances, may provide a simple explanation of the large scale topological structure observed in metabolism~\cite{Jeong:2000}, and it can also be used, like in this work, to revisit the concept of biochemical pathways.

In this paper, we uncover the hidden geometry of the E. coli and human metabolisms and find that their network topologies obey an extremely simple and powerful --metric-based-- probabilistic connectivity law. In particular, given a pair metabolite/reaction separated by a geometric distance $d_{mr}$ in the underlying metric space, the probability of existence of a connection between them is here shown to be a decreasing function of the effective distance $d_{eff} \equiv d_{mr}/(k_r k_m)$, where degrees $k_m$ and $k_r$ count the number of their respective neighboring nodes. The geometric distance $d_{mr}$ --a measure of structural affinity between metabolites and reactions-- is in this way modulated by the product of degrees of the two involved nodes, so that the degree heterogeneity observed in the metabolic network is properly taken into account. Naturally, a key ingredient in our approach concerns the suitable geometry substantiating this distance. We find that a simple one dimensional closed Euclidean space, i.e. a circle, when combined with the network degree heterogeneity is enough to capture the global organization of the network. Using statistical inference techniques, we find angle-based coordinates in this space for the full set of metabolites and reactions, which expose the extraordinary congruency of our model.

As a direct application of the proposed cartographic maps of metabolism, we compare the results of our embedding with
the standard classification of reactions in terms of biochemical pathways. Such a reaction-aggregated analysis reveals rather disparate trends when pathways are characterized in terms of the circle-based localizations of their constituents reactions. Some specific pathways appear concentrated over narrow sectors of polar angles, while more transversal ones are widespread over the circle. This points to a diversity of pathway topologies, with some of them displaying groups of densely interconnected reactions while some others evidencing a much more weakly connected internal structure. Moreover, pathways themselves admit to be linked using the discovered connectivity law. This strategy reveals different levels of cross-talk between pathways, leading to a coarse-grained view of metabolic networks or, in other words, to the build-up of networks of pathways. Such a higher level in the hierarchical organization of metabolic networks advises against the study of pathways as autonomous subsystems and should permit to calibrate more accurately how a pathway-localized perturbation spreads over the entire network.

\section{Results}
\subsection{Embedding algorithm and validation}

\begin{figure}[!ht]
\centering
\includegraphics[width=8.7cm]{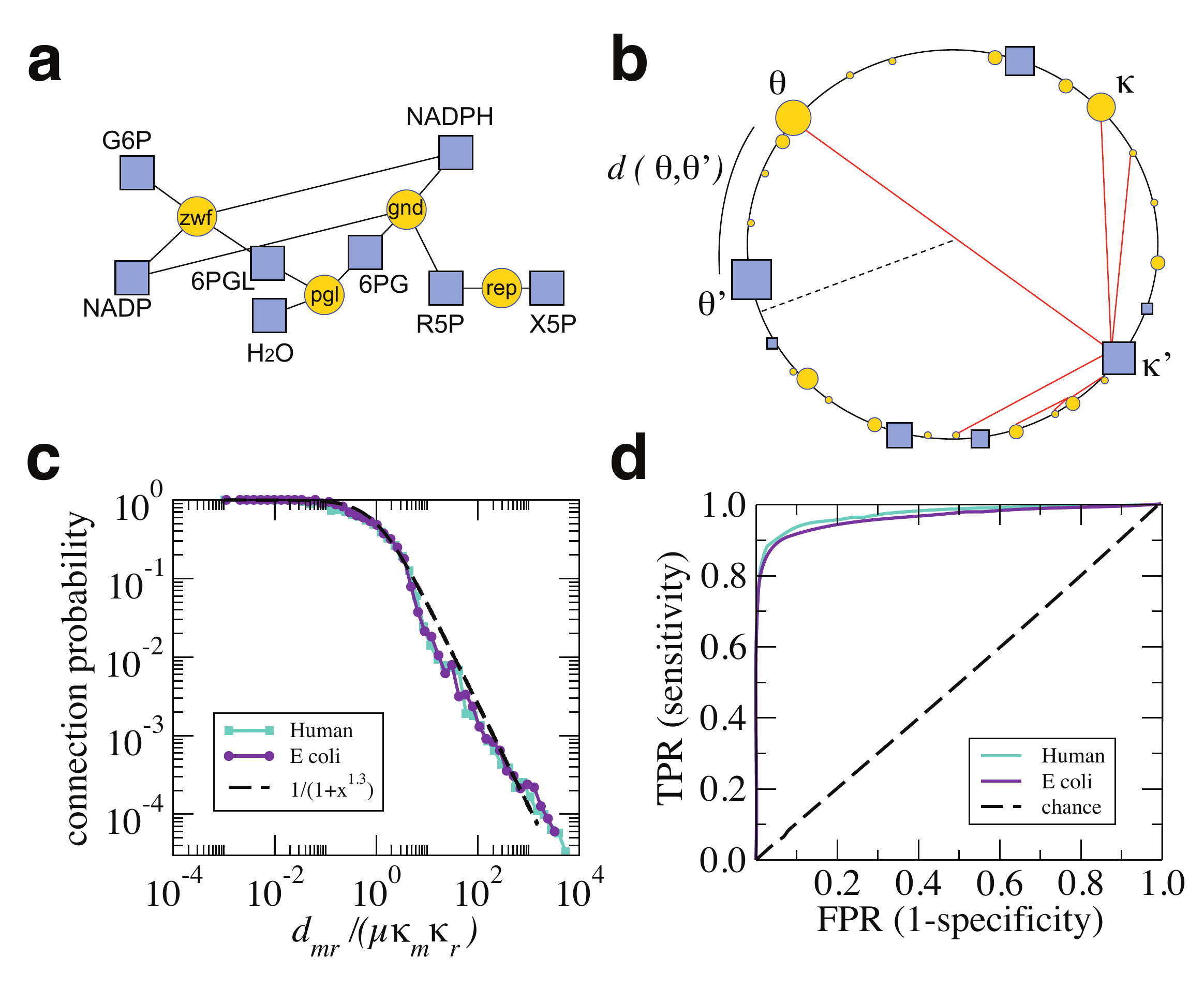}
\caption{
{\bf Model and empirical validation.}
{\bf a}, Bipartite network representation of four coupled stoichiometric equations in the pentose-phosphate pathway of {\it E. coli}. Reaction acronyms stand for the catalyzing enzyme: {\it zwf}, glucose- 6- phosphate dehydrogenase [EC 1.1.1. 49] ; {\it pgl}, 6- phospho-gluconolactonase [EC 3. 1. 1.31] ; {\it gnd}, 6- phosphogluconate dehydrogenase [EC 1.1.1. 43] ; {\it rpe}, ribulose- phosphate 3- epimerase [EC 5. 1.3. 1]. Notice that connections (black lines) are always between reactions (yellow circles) and metabolites (blue squares), metabolites or reactions are never connected among themselves. {\bf b}, A sketch of the $\mathbb{S}^1\times \mathbb{S}^1$ model. Nodes are randomly distributed in the circle and given expected degrees, symbolically represented by the sizes of the nodes. The distance between two nodes is computed as the length of the arc separating the nodes. Due to the peculiar rescaling of distances by degrees in Eq.~(\ref{eq:1}), a node can connect not only to nearby nodes but also to far apart nodes with large degree. {\bf c}, The plot shows a comparison between the empirical connection probability for the E. coli and human metabolisms and the theoretical one given in Eq.~(\ref{eq:2}). The empirical connection probability is computed as the fraction between the number of actual connections at effective distance $d_{mr}/\mu k_m k_r$ and the total number of pairs at the same effective distance.
{\bf d}, The Receiver Operating Characteristic (ROC) curve computed for our model for the E. coli and human metabolisms is shown. To calculate the ROC curves, we rank (from highest to lowest) the connection probabilities given by the model for all possible pairs metabolite/reaction (either present or absent) using the previously inferred coordinates. We then define at each value a threshold probability that allows us to discriminate between positive interactions (those above the threshold) from negative ones (those below the threshold) and to compute the fraction of true positive connections (True Positive Rate TPR) and that of false positive connections (False Positive Rate FPR), with the understanding that a true positive connection is an observed link above the threshold, while a false positive is an non-existing one above the threshold.
}
\label{fig:1}
\end{figure}

A simple abstraction of a given metabolism is given by its bipartite network representation. This amounts to consider metabolites and reactions as belonging to different subsets of nodes, with metabolites (irrespectively considered as reactants and products) linked to all reactions they take part in, and thus avoiding connections between nodes of the same kind, see Fig.~\ref{fig:1}a. The first step towards mapping this network consists in defining a geometric model that can advantageously represent it. The simplest realization of a one-dimensional homogeneous and isotropic closed metric space that can globally embed a network is a circle of radius $R$. Nodes, in our case metabolites and reactions separately, are distributed on it according to specific angular coordinates to be determined.
The whole strategy to find these coordinates rests on a precise definition of the interactions between nodes in terms of their ring-based distances. We prescribe a connection probability between a reaction $r$ and a metabolite $m$, with respective bipartite degrees $k_r$ and $k_m$ and separated by a distance $d_{mr}$ on the circle ($d_{mr} = R \Delta \theta_{mr}$, $\Delta \theta_{mr}$ being the angular separation between metabolite and reaction) to be a decreasing function of such distance rescaled by the product of node degrees~\cite{Serrano:2008a},
\begin{equation}
\mbox{Prob\{m is connected to r\}}\equiv
p\left(\frac{d_{mr}}{k_m k_r}\right).
\label{eq:1}
\end{equation}
It is worth-stressing that this is the central and unique law underlying the whole formalism. Notice that this choice is particularly suggestive since by identifying the node degree as a measure of its mass, this interaction mimics the Newtonian form of gravitational interaction. More precisely, the explicit form for the above interaction reads
\begin{equation}
\label{eq:2}
p\left(\frac{d_{mr}}{k_m k_r}\right) = \frac{1}{1+(d_{mr}/\mu k_m k_r)^{\beta}}.
\end{equation}
This particular prescription combines, in a simple way, the classical network topological concept of node degrees with the newly introduced notion of geometric distance.  All in all, this functional form expresses an intuitive view, i.e. closer nodes in the metric space are more likely to be linked, while nodes with higher  degrees sustain farther reaching connections regardless of their distances. Figure~\ref{fig:1}b shows a visual sketch summarizing the basic trends of the bipartite formalism just outlined. We refer to it with the notation $\mathbb{S}^1 \times \mathbb{S}^1$, see {\it Methods}. Besides, this model gives rise to a maximum-entropy ensemble of graphs that are therefore maximally random given their specific constraints~\cite{PhysRevE.78.015101,PhysRevE.82.036106}. Finally, parameters $\mu$ and $\beta$ are consistently determined to reproduce the statistical properties of the original network. Parameter $\mu$ fixes the total number of edges, whereas $\beta$ controls clustering, i.e, a measure of short-range loops, see Appendix B.

To infer the angle-based coordinates for metabolites and reactions in the ring we use a two-step procedure. Starting from the original bipartite network, we first perform a one-mode projection over the set of metabolites by connecting two metabolites whenever they participate in the same reaction. We then circle-embed such a unipartite metabolites network applying the unipartite version of the formalism as described earlier~\cite{Boguna:2010fk}. Finally, using this partial allocation as an initial fixed template, we complete the embedding of the reactions by invoking a maximum likelihood inference strategy (the detailed description of the embedding algorithm and the coordinates of metabolites and reactions are fully reported in Appendix C).

We apply our formalism to the iAF1260 version of the K12 MG1655 strain of E. coli metabolism~\cite{Feist:2007} and to human cell metabolism~\cite{Duarte:2007}, both provided in the BiGG database~\cite{Schellenberger:2010, Bigg}, see Appendix B. Before presenting the embedding for these metabolic networks, we comment on the validation of the proposed mapping procedure. We first perform a direct calibration which amounts to compare the set of observed metabolite-reaction connection probabilities in the original reconstructions with the theoretical connection probability given by Eq.~(\ref{eq:2}). Explicit results are presented in Fig.~\ref{fig:1}c, both for E. coli and human metabolisms. Beyond the striking agreement between observed and predicted connections, it is worth noticing that the two analyzed networks are perfectly represented with the same $\beta$ exponent fitted to a value $\beta =1.3$. We also check the discrimination power of our algorithm by computing the Receiver Operating Characteristics (ROC) curve of our model~\cite{Fawcett2006861}, which compares the true positive rate (TPR) vs. the false positive rate (FPR) and informs about how good is our method at correctly discern real links. Results are shown in Fig.~\ref{fig:1}d.
When representing the TPR in front of the FPR, a totally random guess would result in a straight line along the diagonal. In contrast, the ROC curve of our model lies far above the diagonal, which indicates a remarkable discrimination power. A convenient summary statistic can be defined as the area under the ROC curve (AUC statistic), which represents the probability that a randomly chosen observed link in the network has a higher probability of existence according to the model than a randomly chosen non-existing one. This statistic ranges in the interval $[0.5,1]$, being $\mbox{AUC}=0.5$ a random prediction and $\mbox{AUC}=1$ a perfect prediction. In our case, values are $\mbox{AUC}=0.96$ for E. coli and $\mbox{AUC}=0.97$ for human metabolism. Both validation tests confirm that our model adjusts nearly perfectly to the real data.

\begin{figure*}[!ht]
\centering
\includegraphics[width=17cm]{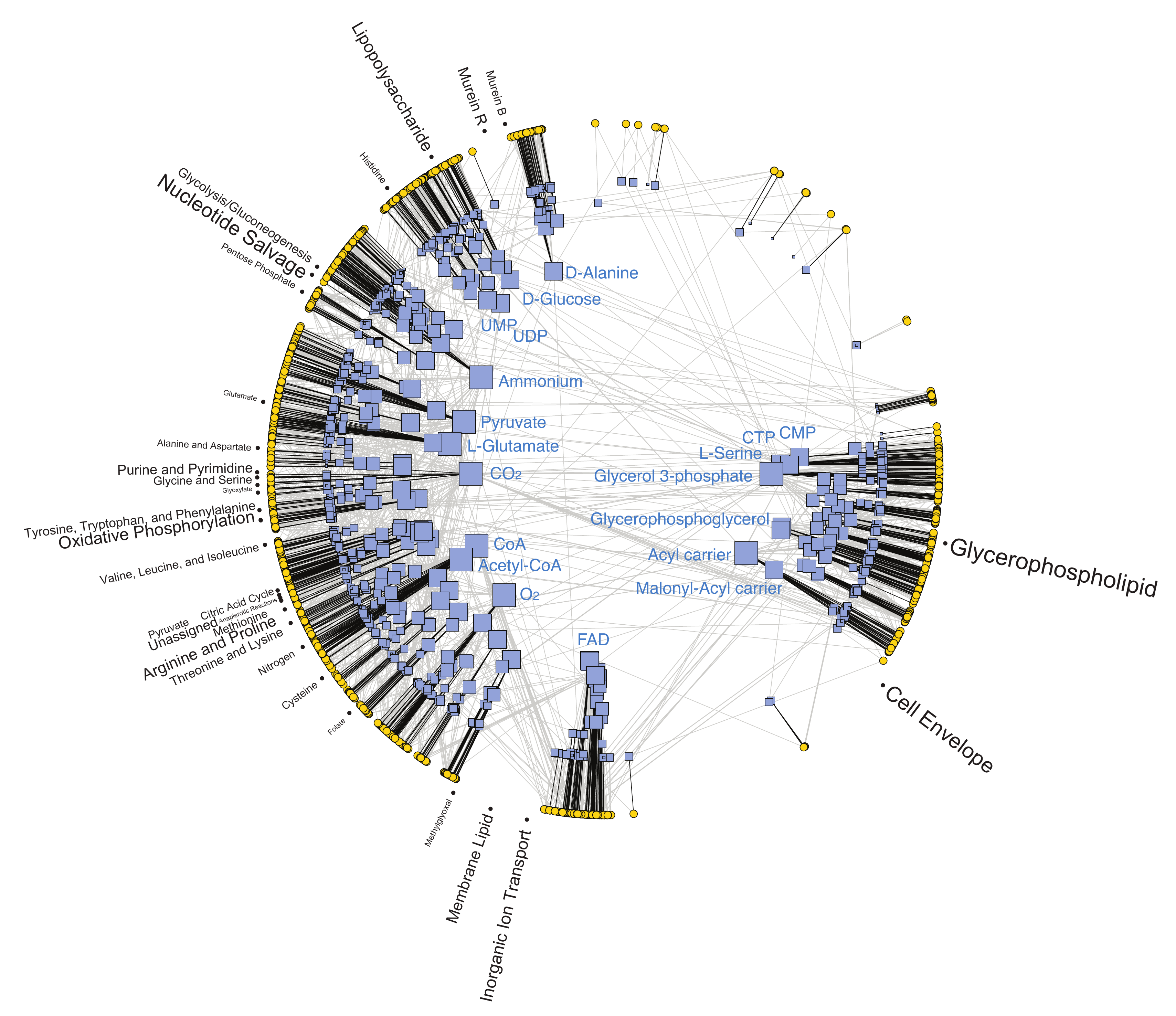}
\caption{{\bf Global geometric map of E. coli's metabolism.} Angular distribution of reactions and metabolites inferred by the method. Yellow circles represent reactions whereas blue squares are metabolites. For each metabolite, the symbol size is proportional to the logarithm of the degree and radially placed according to the expression $r=R-2\ln{k_m}$. Black (grey) connections are those that according to the model have a probability of existence larger (smaller) than $0.5$. The names of the different pathways, radially-written, are located at the average angular position of all the reactions belonging to a given pathway, and the font size is proportional to the logarithm of the number of reactions in the pathway. Notice that we do not represent transversal pathways and that some pathways seem to be located in empty regions (e.g. Inorganic Ion Transport). This is due to the fact that some pathways display bimodal or multi-peaked distributions so that the average appears in between the peaks, see Fig.~\ref{fig:3} and Table I in Appendix D.
}
\label{fig:2}
\end{figure*}
Figure~\ref{fig:2} shows the embedding representation of the E. coli metabolism (the mapping of the human metabolism is provided in Appendix F). For the sake of clarity, metabolites are displaced towards the center of the circle by an amount proportional to their degree so that hub metabolites are close to the center of the disk whereas low degree ones are placed in the periphery. The distribution over the circle is far from being uniform as it could be naively expected. Indeed, this is a distinctive signature of the delicate structural organization of metabolic networks. In particular, different levels of aggregation are readily visible, inasmuch as human settlements are unevenly distributed in population maps. Simultaneously with densely occupied areas, empty regions are visible and appear irregularly punctuated with occasional metabolite-reaction associations. As a whole, this landscape is an indication of some hierarchical trends existing in the analyzed networks and prompts us to look for eventual higher organizational levels. In this regard, we revise the biochemical concept of pathways, classically understood as chains of step-by-step reactions which transform a principal chemical into another either for immediate use, to propagate metabolic fluxes or for cell storage.  In Fig.~\ref{fig:2}, we identify pathways in the circle by plotting their names at the average angular position of all their constitutive reactions.

\subsection{Pathway localization}
\begin{figure}[!ht]
\centering
\includegraphics[width=8.7cm]{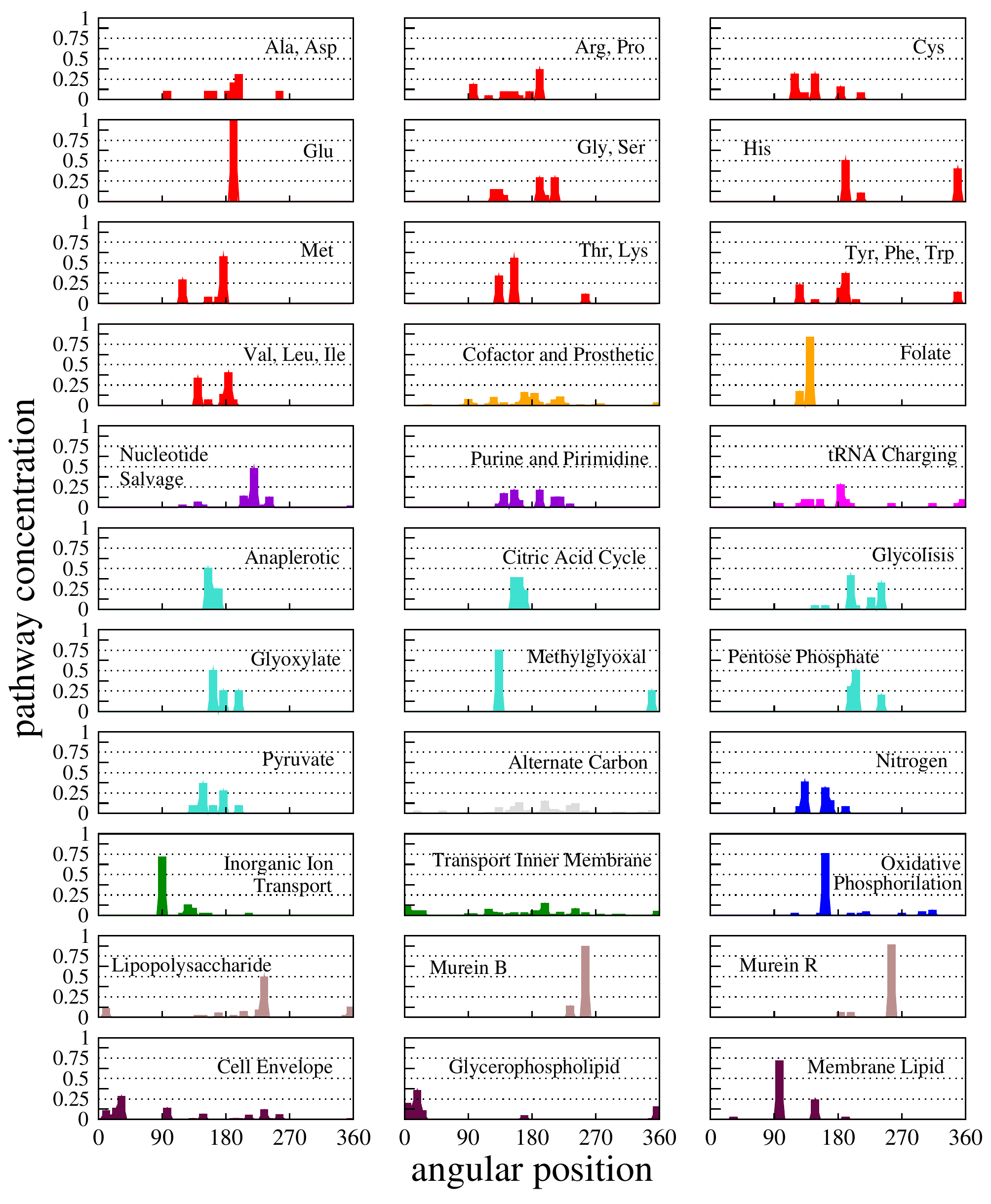}
\caption{{\bf Angular distribution of biological pathways in E. coli.} The whole angular domain $[0,360^o]$ is divided in $50$ bins of $7,2^o$ each and for each bin we compute the fraction of reactions of the pathway in it. Each pathway is shown in a different graph. Different colors indicate different general metabolic classes: red for Amino Acids metabolism (numbering the graphs from left to right and from top to bottom, 1-10), orange for metabolism of Cofactors and Vitamins (11-12), violet for Nucleotide metabolism (13-14), magenta for tRNA charging (15), turquoise for Carbohydrate metabolism (16-22), grey for Alternate Carbon metabolism (23), blue for Energy metabolism (24,27), green for Transport pathways (25-26), brown for Glycan metabolism (28-30), and maroon for Lipid metabolism (31-33). Pathway names have been abbreviated in standard forms whenever possible.
}
\label{fig:3}
\end{figure}

\begin{figure}[!ht]
\centering
\includegraphics[width=8.7cm]{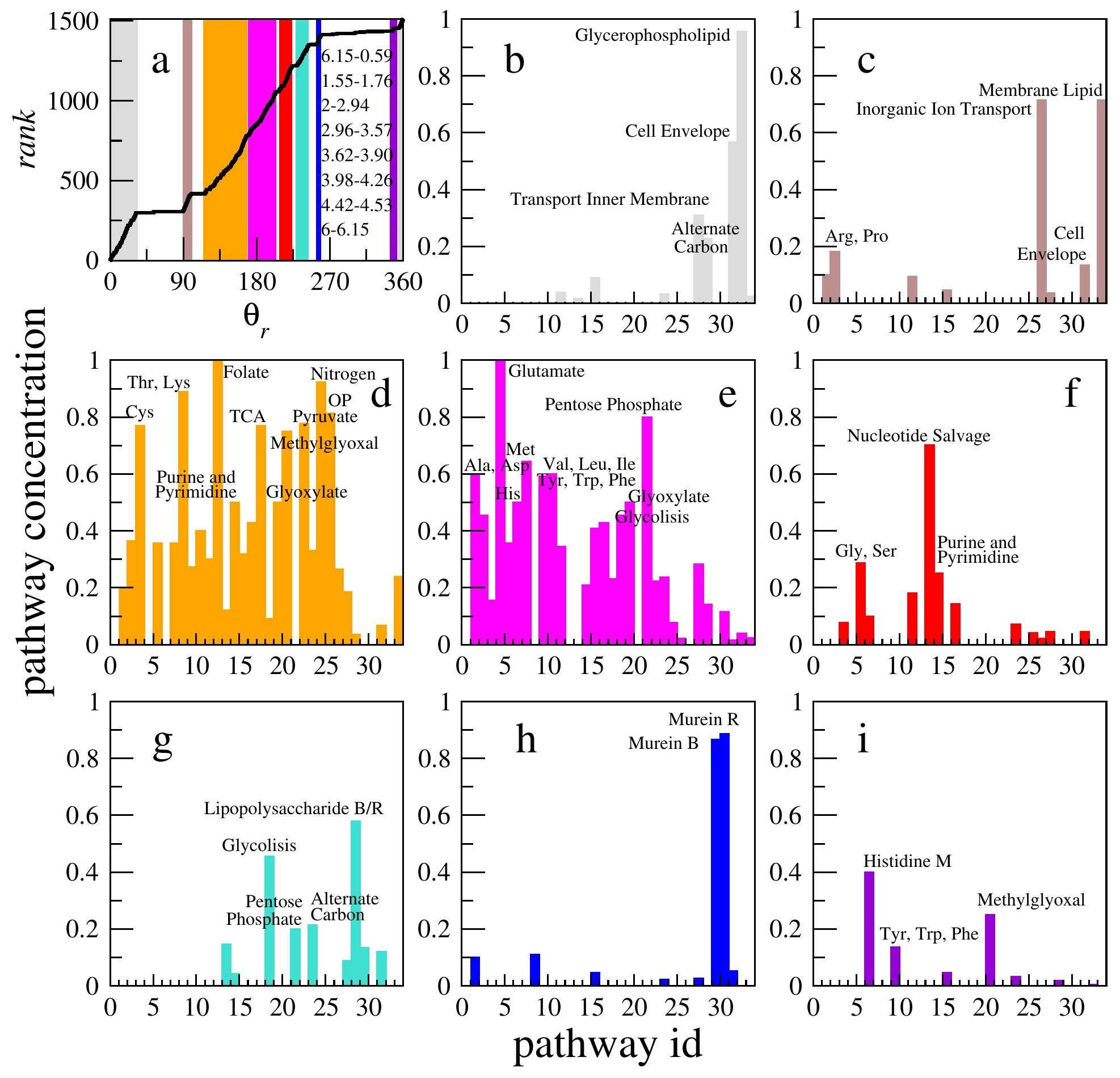}
\caption{{\bf Sector modules for E. coli's metabolism.} Reactions in related functional categories are observed to aggregate in specific regions of the circle. The whole angular domain is divided into eight different angular sectors delimited by void regions in the ranked distribution of reaction angles. This distribution and the angular coordinates defining the sectors are given in the left upper graph of the panel. Each sector is indicated in a different color. The remaining graphs show the pathway concentration, the fraction of reactions of that pathway, in each sector. The higher concentrations in each sector mostly correspond to pathways in related functional categories: S1 and S2 aggregate pathways related to Cell Membrane metabolism (plots b and c), S3 concentrates Central metabolism including Energy and part of the Nucleotide metabolism (plot d), S4 gathers Central metabolism including Amino Acids metabolism (plot e), S5 condenses the remaining Nucleotide metabolism (plot f) and S6 and S7 account for Glycan metabolism (plots g and h), with S6 mixing basically mono and polysaccharide related pathways and pathways related to murein, a polymer that forms the cell wall, well separated in S7. Pathway names have been abbreviated in standard forms whenever possible.
}
\label{fig:4}
\end{figure}
In Figs.~\ref{fig:3} and~\ref{fig:4}, we propose two complementary representations of the metabolic pathways of E. coli as they appear annotated in the BiGG database. In Fig.~\ref{fig:3}, we show the angular distribution on the ring of the whole list of pathways (up to 33, plus an {\it Unassigned} category of reactions not represented in the figure), evaluated from the circle-based embedding of the reactions they involve. We recognize rather disparate spectra of angular distributions. Strongly localized pathways, e. g. the Folate pathway or Oxidative Phosphorylation, coexist with more distributed ones. The latter can adopt either a discrete bimodal, a multi-peaked form, e. g. the Histidine and Glycolisis pathways respectively, or can even transversally spread over the ring closer to a homogeneous distribution. The Alternate Carbon, the Transport Inner Membrane, or the Cofactor and Prosthetic Group pathways are representative examples of this latter category (see Table I in Appendix D for further details). Our method is, therefore, able to discriminate concentrated pathways, consistent with the classical view of modular subsystems, from others which are indeed formed of subunits, and even from those finally responsible of producing or consuming metabolites in turn extensively used by many other pathways.

The embedding of reactions and metabolites in the circle can also be used to aggregate pathways into broader categories. To do so, the embedding circle is first divided into eight different angular sectors delimited by void regions in the ranked distribution of reaction angles, see Fig.~\ref{fig:4}a. The pathway concentration, i.e. the fraction of reactions of that pathway in each sector, is shown in Fig.~\ref{fig:4}b-i. Clearly, there are sectors monopolized by one or at most two pathways --e. g., Murein in Sector 7, Fig.~\ref{fig:4}h--, whereas other sectors are largely shared by many pathways --e.g. , different Aminoacid-based pathways in Sector 4, Fig.~\ref{fig:4}e. In all cases, the higher concentrations in each sector mostly correspond to pathways in related functional categories: Sector 1 and Sector 2 in Fig.~\ref{fig:4}b-c aggregate pathways related to Cell Membrane metabolism, Sector 3 and Sector 4 in Fig.~\ref{fig:4}d-e concentrate Central metabolism, with Sector 3 including Energy and part of the Nucleotide metabolism and Sector 4 including Amino Acid metabolism, Sector 5 in Fig.~\ref{fig:4}f condenses the remaining Nucleotide metabolism, and Sector 6 and Sector 7 in Fig.~\ref{fig:4}g-h account for the Glycan metabolism, with Sector 6 mixing basically mono and polysaccharide related pathways, while the pathways related to murein, a polymer that forms the cell wall, appearing well separated in Sector 7.

Corresponding representations for human metabolism are shown in Fig.~10. The number of pathways is considerably larger but common features to E. coli pathway localization patterns are evidenced in qualitative terms. Pathways can be divided again into different categories according to their angular concentration, with the difference that the general level of pathway localization in human metabolism is higher than in E. coli. The average angular concentration of pathways in human metabolism is $0.82$, as compared to $0.79$ in E. coli (see {\it Methods}) and the average size of maximum peaks in the pathways angular distributions in $0.36$ for E. coli while for human metabolism it is $0.50$. However, the higher level of localization seems to coexist with a higher entanglement of the different families of metabolic reactions, i.e carbon metabolism, lipid metabolism, etc.. Another observation is that transversal pathways in E. coli, like Cofactor and Prosthetic group or Transport, are split into a number of more specialized pathways in human metabolism and, in fact, the category of transversal pathways itself, as defined in E. coli, is here minimally represented.

\subsection{Cross-talk between pathways}
\begin{figure}[!ht]
\centering
\includegraphics[width=8.7cm]{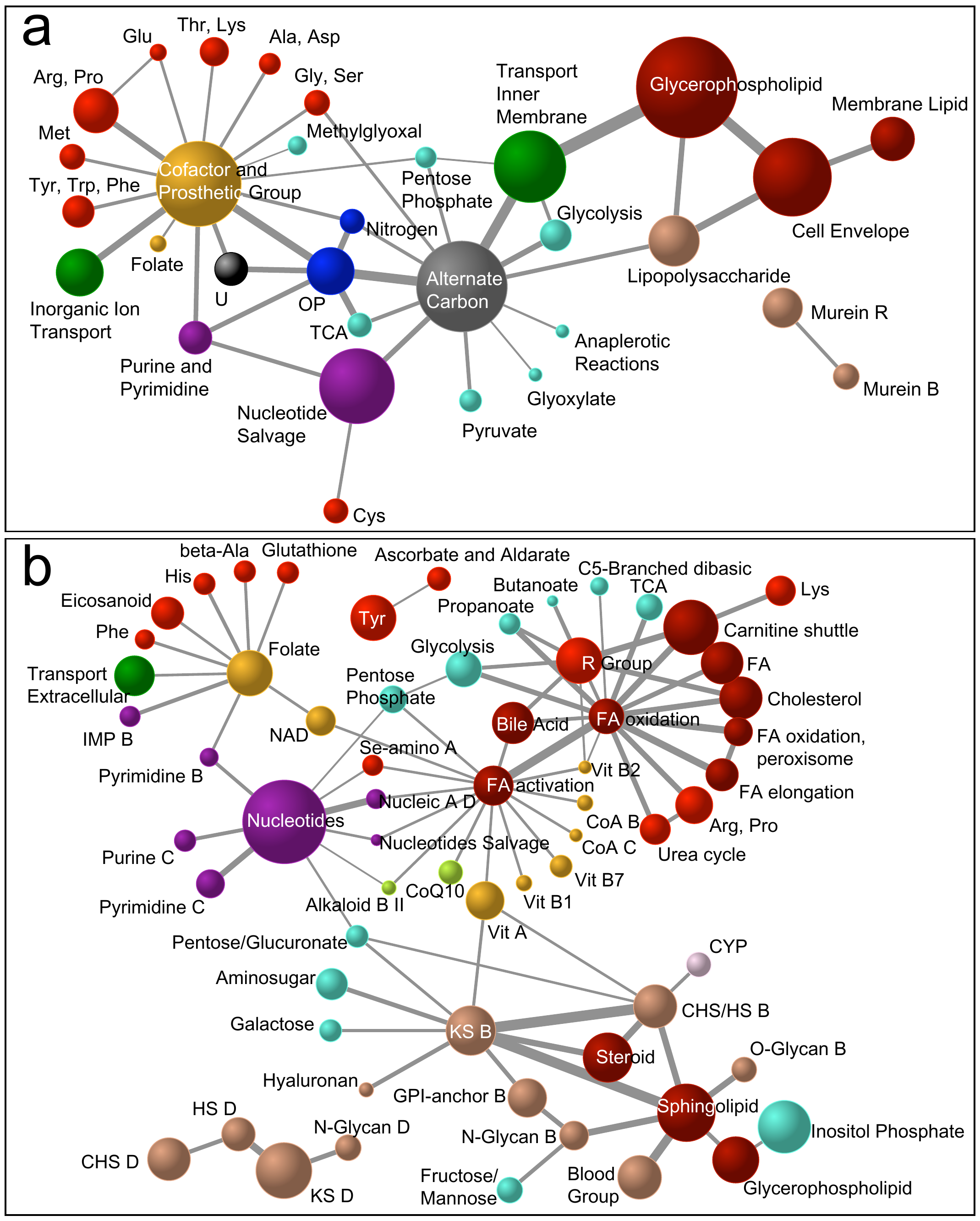}
\caption{{\bf Metabolic backbones displaying pathway's cross-talks inferred from the model}. {\bf a}, Metabolic backbone for E. coli metabolism at the $0.064$ confidence level with $30\%$ of the original total weight, $91\%$ of the original number of pathways, and $9\%$ of the original links. {\bf b}, Metabolic backbone for human cells at the $0.022$ confidence level, with $20\%$ of the original total weight, $69\%$ of the original number of pathways, and $5\%$ of the original links. Different colors indicate different metabolic families as defined in the caption of Fig.~\ref{fig:3}. The area of a circle representing a pathway is proportional to its size in number of reactions.  The weights in the connections are proportional to the intensity of the cross-talk between the pathways. Pathway names have been abbreviated in standard forms whenever possible.
}
\label{fig:5}
\end{figure}
In the first part of the paper, the $\mathbb{S}^1 \times \mathbb{S}^1$ formalism was applied to embed the observed metabolic networks into a circle, enabling to locate the reactions and metabolites related to every specific pathway in a simple one-dimensional geometry. This information can be used to build a higher hierarchical level in the architecture of the metabolic network aimed at quantifying the inter-connectivity between pathways. In turn this allows us to introduce the concept of network of pathways.

Adjacencies between a pair of pathways are computed on the basis of the corresponding lists of reactions in each pathway and the set of metabolites shared by both lists. When the set of overlapping metabolites is not empty, the connection probabilities for the links between pathway reactions and common metabolites that correspond to observed interactions in the network are summed to give an absolute measure of the strenght of the interaction between the pair of pathways. Overlaps between pathways pairs assemble a higher order weighted network where pathways are nodes and links display heterogeneous intensities. However, the resulting network is very dense and needs to be conveniently filtered in order to provide meaningful information about the system. In E. coli, $460$ out of a potential total of $561$ pathways pairs overlap while for human cells $1689$ pathways pairs out of $4278$ have common metabolites. In practice we use a disparity-based threshold~\cite{Serrano:2009b} (see {\it Methods}) that discards links whose intensities are compatible with random fluctuations at some specific significance level. As a result these pathway-based networks provide metabolic backbones i.e., subnetworks of pathways which display the statistically relevant interactions.

As an illustration of the power of the metabolic backbone concept, panels in Fig. 5 reproduce the corresponding constructions for E. coli and human metabolisms. Interestingly, metabolic backbones offer a perspective that reveals functional constraints. Both for E. coli and human metabolism, star-like patterns are particularly neat. In E. coli, transversal pathways act as hub-like structures that interconnect different number of specific and more localized pathways, usually belonging to the same metabolic family. For instance, the Cofactor and Prosthetic Group Biosynthesis pathway connects many of the amino acid pathways to energy or nucleotide metabolism, and Alternate Carbon acts as the main intermediary of many Carbohydrate pathways with the rest of the backbone. Analogously, some pathways in the metabolic backbone of the human cell, like Folate or Fatty Acid Oxidation or Keratan Sulfate Biosynthesis, play a relevant role in providing systems' level connectivity to the network and connect a number of other specific pathways.

\section{Discussion}
From a broad perspective, a cartographic representation of complex networks supposes to map the positions of nodes in an underlying geometric space and shares some fundamental problems with traditional geographical cartography on what concerns techniques, generalizations or design: how to represent the topology of the mapped network on the metric space, which characteristics of the network are not relevant to the map's purpose and can be eliminated, how to reduce the complexity of the characteristics that will be mapped, etc.. Despite the difficulties, cartographic maps based on geometrical spaces are crucial to identify dominant nodes, to understand how different subparts of the system, like pathways in our context, relate to each other, to back up more accurate methods of prediction of missing and spurious interactions~\cite{Clauset:2008,Guimera:2009}, or to find optimal transport routes.

In our metabolic maps, the astonishing congruency between the architecture of metabolic networks and the underlying geometry is supported by a biochemical interaction law that, irrespective of the studied organisms, of the nature and complexity of the reactions they account for, or of the different structural label of the metabolites they involve, seems to comply with a simple Newtonian-like form and allows us to make predictions about the probabilities of interaction among sets of metabolites forming reactions. 
Specifically, the sum of the probabilities running over all the metabolites participating in a certain biochemical reaction 
can be interpreted as a topological version of 
the well-known concept of reaction-based affinity, and each summand could thus be identified with the chemical potential of that particular metabolite in relation to its chemical partners in the particular reaction. Our results point to a systems level definition of chemical affinity in terms of network-based probabilities of interaction which depend on the distances in the underlying geometric space and on intrinsic properties of nodes which convert some of them in hubs.

Such probabilistic network-based chemical affinities allow us to recover the established biochemical organization of pathways as connected metabolic families, but at the same time raise new questions claiming for the need of rethinking its classical definition as self-contained units. We find that different pathways may have disparate internal structures, some of them being more modular and conforming better to the classical definition, while subunits pointing to differentiated functionalities can be distinguished in others. We have also unveiled a higher level of systems' level interactions represented by metabolic backbones, defined on the basis of a quantitative cross-talk between pathways. This particular idea advises us against the use of very specific biochemical protocols aimed to single-out particular pathways as they might be prone to underestimate the delicate connections that underlay the net and secure its proper functioning. Such metabolic features are common to human cells and E. coli. However, a comparative study shows that pathways in human metabolism are in general more modular and display less overlap of common metabolites with other pathways. At the same time the different human metabolic families are more entangled and sectors are mode difficult to characterize, a possible signature of a higher functional complexity or merely a side effect of the kind of reconstruction that mixes in a single network reactions that happen in diversely differentiated cells.

Summarizing, in this work we provide cartographic maps of two representative metabolisms that capture their specific complexities, explaining many of their system properties and provide a new perspective on the definition, cross-talk, and hierarchical organization of biochemical pathways. These maps, embedded in a simple geometric space, rely on a probabilistic biochemical connectivity law which emerges from the different physico-chemical forces acting at a molecular level and that naturally conveys a higher interaction likelihood to elements which are closer in the underlying space. Similar maps for other biological networks are expected to be equally congruent and to help to transform data into knowledge and knowledge into understanding, paving the way for new discoveries in systems biology prediction and control.

\begin{acknowledgments}
This work was supported by MICINN Projects Nos.\ FIS2010-21781-C02-02, FIS2006-03525, and BFU2010-21847-C02-02; Generalitat de Catalunya grants No.\ 2009SGR838 and 2009SGR1055; the Ram\'on y Cajal program of the Spanish Ministry of Science; ICREA  Academia prize 2010, funded by the Generalitat de Catalunya.

{\it Note added}. The extension of the $\mathbb{S}^1$ model to bipartite networks was also developed simultaneously and independently by Maksim Kitsak and Dmitri Krioukov~\cite{Kitsak}.
\end{acknowledgments}


\appendix
\section{Methods}
\subsection{Hidden metric spaces and the $\mathbb{S}^1\times \mathbb{S}^1$ model.}

The $\mathbb{S}^1\times \mathbb{S}^1$ model can be used as a network generator as follows:
\begin{enumerate}
\item
$N_m$ metabolites and $N_r$ reactions are homogeneously distributed in a circle of radius $R$. The densities of metabolites and reactions in the circle are $\delta_m=N_m/2\pi R$ and $\delta_r=N_r/2 \pi R$, taken independent of the network size. Without loss of generality, one of them can be set to 1.
\item
Metabolites and reactions are assigned expected degrees $k_m$ and $k_r$, drawn from the probability densities $\rho_m(k_m)$ and $\rho_r(k_r)$, respectively. To model metabolic networks, we use $\rho_m(k_m) \sim k_m^{-\gamma}$ and $\rho_r(k_r)=\delta(k_r-\langle k_r \rangle)$.
\item
Each possible pair metabolite/reaction is visited once and a link is created with probability
\begin{equation}
p(k_m,\theta_m;k_r,\theta_r)=p\left(\frac{d_{mr}}{\mu k_m k_r}\right),
\end{equation}
where $d_{mr}=R \Delta \theta_{mr}$ ($\Delta \theta_{mr}$ is the angular separation) is the distance metabolite/reaction in the circle. Function $p$ can be, {\it a priori}, any integrable function. However, the choice $p(x)=(1+x^{\beta})^{-1}$ generates maximally random networks given the constraints of the model.
\end{enumerate}
See Appendix B for extended details on the $\mathbb{S}^1\times \mathbb{S}^1$ model.

\subsection{Inverse problem}
Given a complex network representation, the inverse problem of embedding the network in the hidden metric space amounts to find the optimal position of every node in that underlying geometry. The optimal coordinates would ensure that, given the specific form of the connection probability in Eq.~(\ref{eq:2}), the model has a maximum probability to reproduce the observed topology. In general terms, the embedding is resolved using statistical inference techniques, basically a maximum likelihood estimation in combination with a Monte Carlo method and a Metropolis-Hasting rule to explore and select possible configurations in the underlying space. More precisely, the likelihood functional is defined as
\begin{equation}
{\cal L}\equiv\prod_{m=1}^{N_m} \prod_{r=1}^{N_r}  \left[p\left(\frac{d_{mr}}{\mu k_m k_r}\right)\right]^{a_{mr}} \left[1-p\left(\frac{d_{mr}}{\mu k_m k_r}\right)\right]^{1-a_{mr}}
\end{equation}
where $a_{mr}$ is the bipartite adjacency matrix of the network, defined as $a_{mr}=1$ if metabolite $m$ participates in reaction $r$ and zero otherwise. The bipartite nature of metabolic networks together with the fact that reactions and metabolites have disparate degree distributions precludes to perform the mapping in a single-step. Rather the embedding into the $\mathbb{S}^1\times \mathbb{S}^1$ space runs in two phases: first the one-mode projection of the metabolic subnetwork is embedded into a $\mathbb{S}^1$ space following the numerical optimization procedures described in~\cite{Boguna:2010fk}, and second the inferred angular coordinates of metabolites are used as an input to adjust the position of each individual reaction in the circle. See Appendix C for a more complete description of the $\mathbb{S}^1\times \mathbb{S}^1$ embedding algorithm.

\subsection{The disparity filter}
To extract the metabolic backbone of cross-talks between pathways we apply the multi-scale disparity filter defined in \cite{Serrano:2009b}. The disparity filter exploits local heterogeneity and correlations among weights in complex weighted network representations to extract the network backbone by considering the relevant edges at all the scales present in the system. It ensures that small nodes in terms of strength ($s_i=\sum_{j\equiv i-neighbors} w_{ji}$, sum of incident weights to node $i$) are not neglected and that the backbone remains connected and does not disaggregate into separate clusters. The methodology preserves interactions with a statistically significant intensity for at least one of the two nodes the edge is incident to. To decide whether a connection is relevant, the filter compares against a null hypothesis which assumes that the local weights associated to a node are uniformly distributed at random. In this way one discounts intensities that could be explained by random fluctuations. The disparity filter produces better results in terms of preserving the maximum number of nodes and weights in the backbone with the minimum number of links as compared to a global threshold filter that selects all the links with weights above a certain value, see Fig.~8 in Appendix E.

\subsection{Average angular position and concentration of pathways.}
To find the average angular position of a given pathway and a measure of its angular concentration (or dispersion), we use the following method. Each reaction $i$ of a given pathway (with $i=1, \cdots, N_p$ reactions in it) is assigned a normalized vector $\vec{r}_i$ pointing to the position of the reaction in a circle or radius 1 using as angular coordinate the one inferred by our method. The average angular
position of the pathway is then defined as the angular coordinate of the average vector $\langle \vec{r} \rangle \equiv \sum_{i=1}^{N_p} \vec{r}_i/N_p$. We use this method to plot the names of the different pathways in Fig.~\ref{fig:2}. The modulus of the average vector $|\langle \vec{r} \rangle|$ is a measure of the angular concentration of the reactions. A value $|\langle \vec{r} \rangle|=1$ means that all reactions in the pathway have the same angular coordinates whereas $|\langle \vec{r} \rangle|=0$ indicates a perfect homogeneous distribution over the circle.

\section{The $\mathbb{S}^1$ model and its extension to bipartite networks}

The $\mathbb{S}^1$ model \cite{Serrano:2008a} is a complex network generator able to generate networks which are, simultaneously, scale-free, small-worlds, and highly clustered, as observed in the majority of real networks. Nodes in this model are distributed in a metric space (in the simplest case a one-dimensional circle) abstracting  (di)similarities among the elements of the network. The $\mathbb{S}^1$ model generates
networks according to the following steps:
\begin{enumerate}
\item
Distribute $N$ nodes uniformly over the circle $\mathbb{S}^1$ of
radius $N/(2\pi)$, so that the node density on the circle is fixed
to $1$.
\item
Assign to all nodes a hidden variable $\kappa$ representing their
expected degrees. To generate scale-free networks, $\kappa$ is drawn
from the power-law distribution
\begin{eqnarray}
\rho(\kappa)&=&\kappa_0^{\gamma-1}(\gamma-1)\kappa^{-\gamma},\quad\kappa\in[\kappa_{0},\infty),\\
\kappa_0&=& \langle k \rangle \frac{\gamma-2}{\gamma-1},\label{eq:kappa0_fixed}
\end{eqnarray}
where $\kappa_0$ is the minimum expected degree, and $\langle k \rangle$
is the network average degree.
\item
Let $\kappa$ and $\kappa'$ be the expected degrees of two nodes
located at distance $d=N\Delta\theta/(2\pi)$ measured over the
circle, where $\Delta\theta$ is the angular distance between the
nodes. Connect each pair of nodes with probability $p(x)$, where the \emph{effective} distance is defined as $d_{eff} \equiv d/(\mu \kappa \kappa')$,
and $\mu$ is a constant fixing the average degree.
\end{enumerate}
The connection probability $p(x)$ can be any integrable function.
Here we chose the Fermi-Dirac distribution
\begin{equation}
p(x)=\frac{1}{1+x^\beta},
\end{equation}
where $\beta$ is a parameter that controls clustering in the
network. With this connection probability, parameter $\mu$ becomes
\begin{equation}
\mu=\frac{\beta}{2\pi \langle k \rangle} \sin{\left[\frac{\pi}{\beta}\right]}.
\end{equation}
The expected degree of a node with hidden variable $\kappa$ is
$\bar{k}(\kappa)=\kappa$ and, therefore, the degree distribution
scales as $P(k)\sim k^{-\gamma}$ for large $k$. Notice that this is the reason why in the main text we use degrees instead of expected degrees.

\subsection{The $\mathbb{S}^1\times \mathbb{S}^1$ model}

The $\mathbb{S}^1$ model can be extended to bipartite networks as follows: 
\begin{enumerate}
\item
$N_m$ metabolites and $N_r$ reactions are homogeneously distributed on a circle of radius $R$. The density of metabolites and reactions over the circle are then $\delta_m=N_m/2\pi R$ and $\delta_r=N_r/2 \pi R$. These two densities remain constant in the thermodynamic limit so that the radius of the circle is proportional to the number of metabolites or reactions. 
\item
Each metabolite is assigned a hidden variable $\kappa_m$ and each reaction a hidden variable $\kappa_r$. These random variables follow probability densities $\rho_m(\kappa_m)$ and $\rho_r(\kappa_r)$, respectively.
\item
The connection probability between a reaction with hidden variable $\kappa_r$ and a metabolite with hidden variable $\kappa_r$ separated by a distance $d_{mr}=R \Delta \theta_{mr}$ ($\Delta \theta_{mr}$ being the angular separation) is given by
\begin{equation}
p(\kappa_m,\theta_m;\kappa_r,\theta_r)=p\left(\frac{d_{mr}}{\mu \kappa_m \kappa_r}\right),
\end{equation}
which can be any integrable function.
\end{enumerate}

Using the formalism developed in \cite{Boguna:2003b}, we compute the average degree of a metabolite with hidden variable $\kappa_m$ (notice that since the angular distribution is homogeneous, this quantity does not depend on the angular coordinate of the metabolite and so we chose one that is at $\theta_m=0$) as
\begin{equation}
\bar{k}_m(\kappa_m)=N_r \int d\kappa_r \rho_r(\kappa_r) \frac{1}{2\pi} \int_{-\pi}^{\pi} d \theta p\left(\frac{|\theta| R}{\mu \kappa_m \kappa_r}\right).
\end{equation}
Analogously, the average degree of a reaction with hidden variable $\kappa_r$ is
\begin{equation}
\bar{k}_r(\kappa_r)=N_m \int d\kappa_m \rho_r(\kappa_m) \frac{1}{2\pi} \int_{-\pi}^{\pi} d \theta p\left(\frac{|\theta| R}{\mu \kappa_m \kappa_r}\right).
\end{equation}
By doing the change of variables $x=\frac{\theta R}{\mu \kappa_m \kappa_r}$ and taking the thermodynamic limit $R \rightarrow \infty$, we can write
\begin{equation}
\bar{k}_m(\kappa_m)=2 \mu \delta_r I \langle \kappa_r \rangle \kappa_m,
\label{eq:1}
\end{equation}
\begin{equation}
\bar{k}_r(\kappa_r)=2 \mu \delta_m I \langle \kappa_m \rangle \kappa_r
\label{eq:2},
\end{equation}
where $I=\int_0^{\infty} dx p(x)$. By taking the average again
\begin{equation}
\langle k_m \rangle=2 \mu \delta_r I \langle \kappa_r \rangle \langle \kappa_m \rangle,
\label{eq:3}
\end{equation}
\begin{equation}
\langle k_r \rangle=2 \mu \delta_m I \langle \kappa_m \rangle \langle \kappa_r \rangle.
\label{eq:4}
\end{equation}
We immediately see that the following relation holds
\begin{equation}
\frac{\langle k_m \rangle}{\langle k_r \rangle}=\frac{\delta_r}{\delta_m}=\frac{N_r}{N_m}.
\label{eq:5}
\end{equation}
In terms of the average degrees, parameter $\mu$ takes the form
\begin{equation}
\mu=\frac{\langle k_m \rangle}{2 \delta_r I \langle \kappa_r \rangle \langle \kappa_m \rangle}=\frac{\langle k_r \rangle}{2 \delta_m I \langle \kappa_r \rangle \langle \kappa_m \rangle}
\label{eq:6}
\end{equation}
and, therefore, Eqs. (\ref{eq:1}) and (\ref{eq:2}) can be rewritten as
\begin{equation}
\bar{k}_m(\kappa_m)=\frac{\langle k_m \rangle}{\langle \kappa_m \rangle} \kappa_m
\end{equation}
\begin{equation}
\bar{k}_r(\kappa_r)=\frac{\langle k_r \rangle}{\langle \kappa_r \rangle} \kappa_r
\end{equation}
We always have the freedom to chose the averages of the hidden variables $\kappa_m$ and $\kappa_r$ to coincide with the actual averages of the observable variables $k_m$ and $k_r$, that is, $\langle k_m \rangle=\langle \kappa_m \rangle$ and $\langle k_r \rangle=\langle \kappa_r \rangle$. In this case we can write
\begin{equation}
\bar{k}_m(\kappa_m)=\kappa_m \mbox{\hspace{0.5cm}   and \hspace{0.5cm} } \bar{k}_r(\kappa_r)= \kappa_r
\label{eq:7}
\end{equation}
with parameter $\mu$
\begin{equation}
\mu=\frac{1}{2 \delta_r I \langle \kappa_r \rangle}=\frac{1}{2 \delta_m I \langle \kappa_m \rangle}
\label{eq:8}.
\end{equation}
This is the choice that we shall follow in the rest of the text. The degree distributions can now be easily written as
\begin{equation}
P_m(k_m)=\int d\kappa_m \rho_m(\kappa_m) \frac{1}{k_m!}  \kappa_m^{k_m} e^{-\kappa_m}
\label{eq:9}
\end{equation}
\begin{equation}
P_r(k_r)=\int d\kappa_r \rho_r(\kappa_r) \frac{1}{k_r!}  \kappa_r^{k_r} e^{-\kappa_r}
\label{eq:10}
\end{equation}

\subsection{Specific model for metabolic networks}

In the case of metabolic networks, the distribution of metabolites' degrees is a power law with exponent $\gamma \approx 2.6$ and the distribution of reactions' degrees is Poisson-like. We can generate this type of network by chosing
\begin{equation}
\rho_m(\kappa_m)= (\gamma-1) \kappa_{m,0}^{\gamma-1} \kappa_m^{-\gamma} \mbox{  with  } \kappa_m\ge \kappa_{m,0}=\frac{\gamma-2}{\gamma-1} \langle \kappa_m \rangle
\label{eq:20}
\end{equation}
and
\begin{equation}
\rho_r(\kappa_r)=\delta(\kappa_r-\langle \kappa_r \rangle).
\end{equation}
Reaction degrees are then Poisson distributed, that is,
\begin{equation}
P_r(k_r)= \frac{1}{k_r!} \langle \kappa_r \rangle^{k_r} e^{-\langle \kappa_r \rangle}
\end{equation}
whereas the degree distribution of metabolites is
\begin{equation}
P_m(k_m)=(\gamma-1) \kappa_{m,0}^{\gamma-1} \frac{\Gamma(k_m+1-\gamma,\kappa_{m,0})}{k_m!}
\end{equation}
We also chose the connection probability
\begin{equation}
p(x)=\frac{1}{1+x^{\beta}}
\label{p(x)}
\end{equation}
so that the integral $I=\pi/(\beta \sin{(\pi/\beta))}$. We can also chose $\delta_m=1$ without loss of generality. Therefore, the number of relevant (free) parameters of the model are $\langle \kappa_r \rangle$, $\langle \kappa_m \rangle$, $\beta$, and $\gamma$.

\subsection{Parameters estimation and finite size effects}

All results in the previous section are strictly true in the thermodynamic limit. In finite size networks, some of the expressions have to be corrected by size dependent factors as we will show below. Besides, there is an extra complication due to the fact that this model can generate nodes with zero degree, which are never observed in a real network. 

Suppose we are given a real network with $N_m^{obs}$ metabolites and $N_r^{obs}$ reactions and average degrees $\langle k_m \rangle^{obs}$ and $\langle k_r \rangle^{obs}$ with exponent $\gamma$. We now want to estimate the values of $\langle \kappa_r \rangle$, $\langle \kappa_m \rangle$, $N_m$ and $N_r$ in our model. The first complication arises because in our model, out of the $N_m$ nodes, there is a fraction $P_m(0)N_m$ nodes with zero degree that cannot be observed. Therefore, if we observe $N_m^{obs}$ metabolites, the best estimation of $N_m$ is
\begin{equation}
N_m=\frac{N_m^{obs}}{1-P_m(0)}
\end{equation}
and, analogously
\begin{equation}
N_r=\frac{N_r^{obs}}{1-P_r(0)}
\end{equation}

The second complication is due to the fact that the average degree of a power law distribution strongly depends on the maximum degree observed in the sample. For instance, in the case of our $\rho_m(\kappa_m)=(\gamma-1) \kappa_{m,0}^{\gamma-1} \kappa_m^{-\gamma}$, if the sample is finite, the distribution is truncated at a certain value $\kappa_{m,c}$ that, typically, increases with the size of the sample. If we compute the average of $\rho_m(\kappa_m)$ but only up to the maximum $\kappa_m$ observed, we have
\begin{equation}
\langle \kappa_m (\kappa_{m,c}) \rangle = (\gamma-1) \kappa_{m,0}^{\gamma-1}\int_{\kappa_{m,0}}^{\kappa_{m,c}} \kappa_m^{1-\gamma} d\kappa_m
\end{equation}
and so
\begin{equation}
\langle \kappa_m (\kappa_{m,c}) \rangle =\langle \kappa_m \rangle \left(1-\left(\frac{\kappa_{m,0}}{\kappa_{m,c}}\right)^{\gamma-2} \right)
\end{equation}
Notice that this large parenthesis converges to 1 in the thermodynamic limit but for $\gamma \approx 2$ it can be fairly large even for large systems. Let us call this factor $\alpha(\kappa_{m,c})$, that is,
\begin{equation}
\alpha(\kappa_{m,c}) \equiv \left(1-\left(\frac{\kappa_{m,0}}{\kappa_{m,c}}\right)^{\gamma-2} \right)
\label{eq:29}
\end{equation}

Now we need to keep track of the finite size effects from the very beginning. This means that we have to correct Eqs. (\ref{eq:1}) and (\ref{eq:2}) as follows
\begin{equation}
\bar{k}_m(\kappa_m;\kappa_{mc})=\kappa_m
\end{equation} 
\begin{equation}
\bar{k}_r(\kappa_r;\kappa_{mc})=\alpha(\kappa_{m,c}) \kappa_r
\end{equation} 
and taking averages
\begin{equation}
\langle k_m(\kappa_{mc}) \rangle= \alpha(\kappa_{m,c}) \langle \kappa_m \rangle 
\end{equation} 
\begin{equation}
\langle k_r(\kappa_{mc}) \rangle =\alpha(\kappa_{m,c}) \langle \kappa_r \rangle
\end{equation} 
Notice that, to write these set of equations we have used that variable $\kappa_r$ is not power law distributed.

Still, this average $\langle k_m (\kappa_{m,c}) \rangle$ cannot be directly identified with the measured average degree because it also accounts for nodes of zero degree. To correct for this effect, we write
\begin{equation}
\langle k_m \rangle^{obs}=\frac{\langle k_m (\kappa_{m,c}) \rangle}{1-P_m(0)}
\label{eq:34}
\end{equation}
and so
\begin{equation}
\langle \kappa_m \rangle=\frac{1-P_m(0)}{\alpha(\kappa_{m,c})} \langle k_m \rangle^{obs}
\label{eq:35}
\end{equation}
and analogously
\begin{equation}
\langle \kappa_r \rangle=\frac{1-P_r(0)}{\alpha(\kappa_{m,c})} \langle k_r \rangle^{obs}
\label{eq:36}
\end{equation}
with
\begin{equation}
P_m(0)=(\gamma-1) \kappa_{m,0}^{\gamma-1} \Gamma(1-\gamma,\kappa_{m,0})
\label{eq:37}
\end{equation}
\begin{equation}
P_r(0)=e^{-\alpha(\kappa_{m,c}) \langle \kappa_r \rangle}
\label{eq:38}
\end{equation}
\begin{equation}
\kappa_{m,c}=k_m^{max, obs}
\label{eq:39}
\end{equation}
Plugging Eqs. (\ref{eq:20}), (\ref{eq:29}), (\ref{eq:37}), and (\ref{eq:39}) into Eq. (\ref{eq:35}), we obtain a closed equation for $\langle \kappa_m \rangle$ that can be solved numerically. Once this parameter is known, by inserting it into Eqs. (\ref{eq:29}) and (\ref{eq:37}) we obtain the values of $\alpha(\kappa_{m,c})$ and $P_m(0)$. Finally, with the value of $\alpha(\kappa_{m,c})$ and Eqs. (\ref{eq:36}) and (\ref{eq:38}) we get the values of $\langle \kappa_r \rangle$ and $P_r(0)$.
\begin{figure}[t]
\includegraphics[width=8cm]{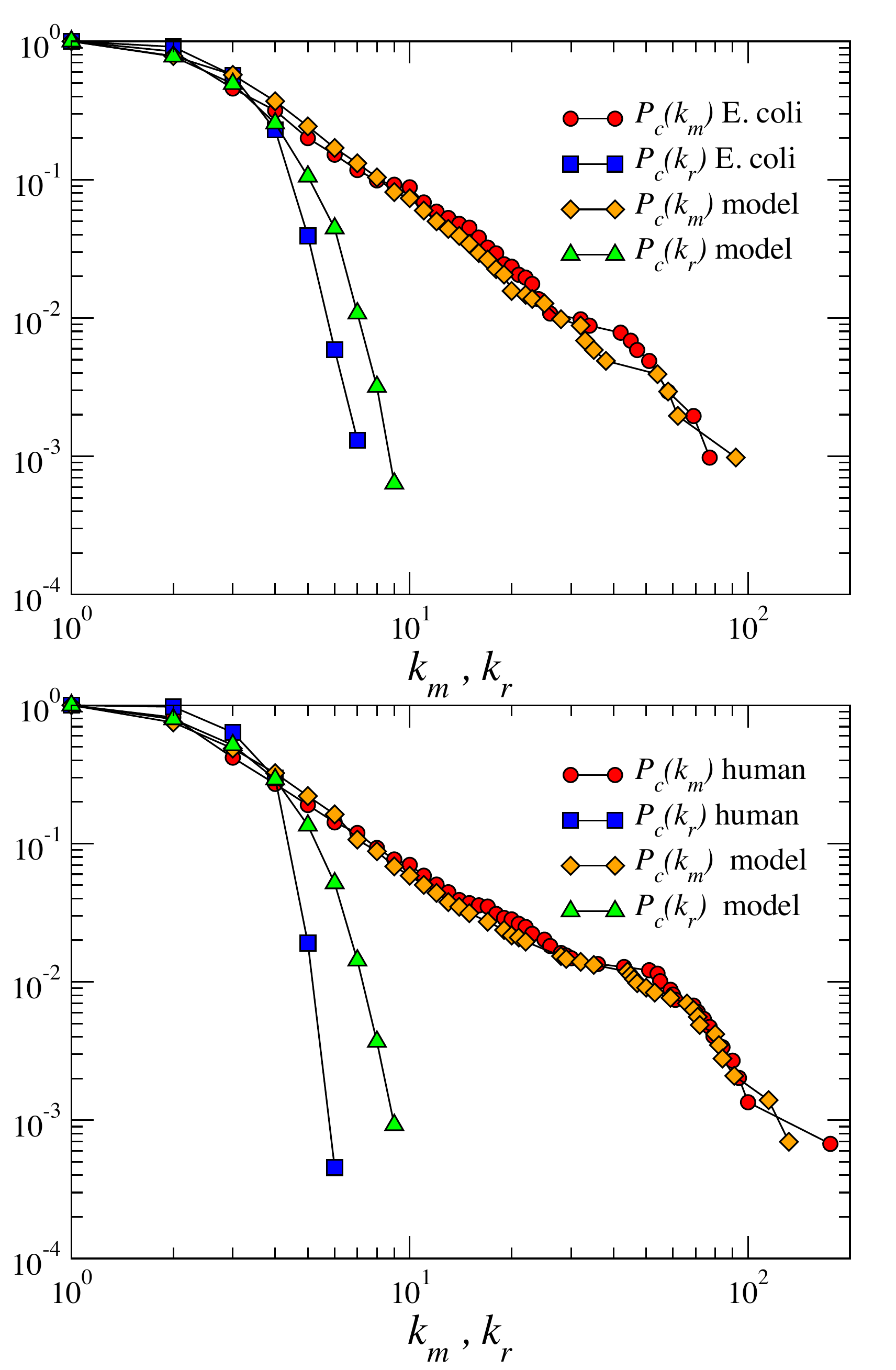}
\caption{{\bf Empirical vs. model degree distributions} Complementary cumulative degree distribution (defined as $P_c(k)=\sum_{k'=k} P(k)$) of metabolites and reactions degrees for the E. coli and human metabolism as compared to two networks generated with the model using the parameters in the text.}
\label{fig:validation0}
\end{figure}

\subsection{Parameters of the real metabolisms}
Using information from the BiGG database \cite{Schellenberger:2010,Bigg}, we build bipartite metabolic network representations of the two analyzed metabolisms, E. coli and human, avoiding reactions that do not involve direct chemical transformations, such as diffusion and exchange reactions. The bipartite representation differentiates two subsets of nodes, metabolites and reactions, mutually interconnected through unweighted and undirected links, without self-loops or dead end reactions. In particular, we analyze the {\it i}AF1260 version of the K12 MG1655 strain of the metabolism of E. coli \cite{Feist:2007}, and the existing annotated list for human metabolism \cite{Duarte:2007}. For the sake of simplicity and to enhance the resolution of the applied algorithm, currency metabolites are eliminated, altogether with a few isolated reaction-metabolite pairs and reaction-metabolite-reaction triplets. For E. coli, this leads to a final set of $1512$ reactions and $1010$ metabolites while human metabolism is nearly $3/2$ larger, with $2201$ reactions and $1482$ metabolites. Characteristic power-law degree distributions for metabolites are readily identified in both organisms, with exponents that are rather similar, respectively $2.65$ for E. coli and $2.55$ for human. Reactions, meanwhile, conform to Poisson-like distributions, whose average values are $2.77$ and $2.93$ respectively. We used the software ``Pajek'' to elaborate all network representations in this paper figures'. 
\begin{itemize}
\item
To find the parameters of the E. coli metabolic network, we use a version of the network where different isomers are considered as different metabolites. Further, we remove the following currency metabolites: h-841, h2o-694, atp-338, pi-308, adp-260, ppi-129, nad-115, nadh-109, amo-85, nadp-83, nadph-81. Ten isolated metabolite-reaction pairs and six isolated reaction-metabolite-reaction triplets have also been removed. For this network, we measure $N_m^{obs}=1010$, $N_r^{obs}=1512$, $\langle k_m \rangle^{obs}=4.15$, and $\langle k_r \rangle^{obs}=2.77$. Using the formalism described in the previous section, we obtain the following estimation of the parameters: $\langle \kappa_m \rangle =4.06$, $\langle \kappa_r \rangle=2.65$, $N_m=1123$, and $N_r=1720$, and $R=N_m/2\pi=178.7$. 
\item
In the case of the Human metabolism, the removed currency metabolites are: h-1250, h2o-916, atp-309, coa-277, pi-240, adp-237, o2-212, nadp-210, nadph-207, nad-202, nadh-195, ppi-114. Three isolated metabolite-reaction pairs have also been removed. We then measure $N_m^{obs}=1482$, $N_r^{obs}=2201$, $\langle k_m \rangle^{obs}=4.34$, and $\langle k_r \rangle^{obs}=2.93$, which leads to the following estimation of the parameters: $\langle \kappa_m \rangle =4.22$, $\langle \kappa_r \rangle=2.73$, $N_m=1646$, and $N_r=2326$, and $R=N_m/2\pi=235.9$. 
\end{itemize}

In Fig. \ref{fig:validation0}, we show the degree distributions for both E. coli and human metabolisms and compare them with those corresponding to networks generated by the $\mathbb{S}^1 \times \mathbb{S}^1$ model. The exponent $\beta$ takes the value $\beta=1.3$ in both networks. The agreement between the model and the real metabolic networks is very good for metabolites. However, the model overestimates the probability of reactions involving five or more metabolites.

\section{Embedding algorithm and validation on $\mathbb{S}^1 \times \mathbb{S}^1$ synthetic networks}

Once the parameters $\langle \kappa_r \rangle$, $\langle \kappa_m \rangle$, $\beta$, and $\gamma$ are estimated, we perform the embedding of the bipartite network to infer the angular coordinates of metabolites and reactions. Let $\mathbb{A} \equiv (a_{ij})_{N_m \times N_r}$, $i=1,\cdots,N_m$, $j=1,\cdots,N_r$, be the adjacency matrix of the network, defined as $a_{ij}=1$ if metabolite $i$ participate in reaction $j$ and zero otherwise (in the rest of the text, symbol $i$ is reserved to enumerate metabolites and symbol $j$ to reactions). Our goal is to find the set of coordinates
$\{\kappa_{m,i},\theta_{m,i}, \theta_{r,j}\}$ that best match the
$\mathbb{S}^1 \times \mathbb{S}^1$ model in a statistical sense. To this end, we
use maximum likelihood estimation (MLE) techniques. Let us compute
the posterior probability, or likelihood, that a network given by
its adjacency matrix $\mathbb{A}$ is generated by the
$\mathbb{S}^1\times \mathbb{S}^1$ model, ${\cal
L}(\mathbb{A})$. This probability is
\begin{eqnarray}
{\cal L}(\mathbb{A})=\int \cdots \int {\cal L}(\mathbb{A},\{\kappa_{m,i},\theta_{m,i}, \theta_{r,j}\}) \cdot \nonumber \\ \cdot \prod_{i=1}^{N_m} d\theta_{m,i} d \kappa_{m,i} \prod_{j=1}^{N_r} d\theta_{r,j} ,
\end{eqnarray}
where function ${\cal
L}(\mathbb{A},\{\kappa_{m,i},\theta_{m,i}, \theta_{r,j}\})$ within the
integral is the joint probability that the model generates the
adjacency matrix $\mathbb{A}$ and the set of hidden variables
$\{\kappa_{m,i},\theta_{m,i}, \theta_{r,j}\}$ simultaneously. Using Bayes' rule, we can compute the likelihood
that nodes' coordinates take particular values
$\{\kappa_{m,i},\theta_{m,i}, \theta_{r,j}\}$ given the observed
adjacency matrix $\mathbb{A}$. This probability is simply given by
\begin{eqnarray}
{\cal L}(\{\kappa_{m,i},\theta_{m,i}, \theta_{r,j}\}| \mathbb{A})=\frac{{\cal L}(\mathbb{A},\{\kappa_{m,i},\theta_{m,i}, \theta_{r,j}\})}{{\cal L}(\mathbb{A})} = \nonumber \\
\frac{\mbox{Prob}(\{\kappa_{m,i},\theta_{m,i}, \theta_{r,j}\}){\cal L}(\mathbb{A}|\{\kappa_{m,i},\theta_{m,i}, \theta_{r,j}\})}{{\cal L}(\mathbb{A})},
\label{likelihood}
\end{eqnarray}
where
\begin{equation}
\mbox{Prob}(\{\kappa_{m,i},\theta_{m,i}, \theta_{r,j}\})=\frac{1}{(2\pi)^{N_m+N_r}}\prod_{i=1}^{N_m} \rho_m(\kappa_{m,i})
\end{equation}
is the prior probability of the hidden variables given by the model,
\begin{equation}
{\cal L}(\mathbb{A}|\{\kappa_{m,i},\theta_{m,i}, \theta_{r,j}\})=\prod_{i=1}^{N_m} \prod_{j=1}^{N_r} p(x_{ij})^{a_{ij}} [1-p(x_{ij})]^{1-a_{ij}}
\end{equation}
is the likelihood of observing $\mathbb{A}$ if the hidden variables are
$\{\kappa_{m,i},\theta_{m,i}, \theta_{r,j}\}$,
\begin{equation}
x_{ij}=\frac{N_r \Delta \theta_{ij}}{\beta \sin{(\pi/\beta)} \kappa_{m,i}},
\end{equation}
\begin{equation}
\Delta\theta_{ij}=\pi-|\pi-|\theta_{m,i}-\theta_{r,j}||,
\end{equation}
and $p(x)$ is given by Eq. (\ref{p(x)}).

The MLE values of the hidden variables $\{\kappa_{m,i}^{*},\theta_{m,i}^{*}, \theta_{r,j}^{*}\}$
are then those that maximize the likelihood in
Eq.~(\ref{likelihood}) or, equivalently, its logarithm,
\begin{eqnarray}
\ln{{\cal L}(\{\kappa_{m,i},\theta_{m,i}, \theta_{r,j}\}| \mathbb{A})}=
C-\gamma \sum_{i=1}^{N_m} \ln{\kappa_{m,i}}+ \nonumber \\
+\sum_{i=1}^{N_m} \sum_{j=1}^{N_r} \left\{ a_{ij} \ln{p(x_{ij})}+ (1-a_{ij}) \ln{[1-p(x_{ij})]} \right\},
\label{likelihoodfinal}
\end{eqnarray}
where $C$ is independent of the nodes' coordinates $\{\kappa_{m,i},\theta_{m,i}, \theta_{r,j}\}$.

\subsection{MLE for expected metabolites' degrees $\kappa_m$}

The derivative of Eq.~(\ref{likelihoodfinal}) with respect to
expected degree $\kappa_{m,l}$ of metabolite $l$ is
\begin{eqnarray}
\frac{\partial}{\partial \kappa_{m,l}} \ln{{\cal L}(\{\kappa_{m,i},\theta_{m,i}, \theta_{r,j}\}| \mathbb{A})}=\nonumber \\
=-\frac{\gamma}{\kappa_{m,l}}-\frac{\beta}{\kappa_{m,l}} \left( \sum_{j=1}^{N_r} p(x_{lj})-\sum_{j=1}^{N_r} a_{lj}\right).
\end{eqnarray}
The first term within the parenthesis is the expected degree of metabolite
$l$, while the second term is its actual degree $k_{m,l}$. Therefore,
the value $\kappa_{m,l}^*$ that maximizes the likelihood is given by
\begin{equation}
\bar{k}(\kappa_{m,l}^*)=\kappa_{m,l}^*=k_{m,l}-\frac{\gamma}{\beta}.
\end{equation}
Since $\kappa_l^*$ can be smaller than $\kappa_0$ in the last
equation, we set
\begin{equation}\label{kappa*}
\kappa_{m,l}^*=\max{\left(\frac{\gamma-2}{\gamma-1} \langle \kappa_m \rangle,k_{m,l}-\frac{\gamma}{\beta}\right)}.
\end{equation}

\addcontentsline{toc}{subsubsection}{MLE for angular coordinates $\theta$}
\vspace{1cm}
\centerline{\bf MLE for angular coordinates $\theta$}\
\begin{figure*}[t]
\includegraphics[width=15cm]{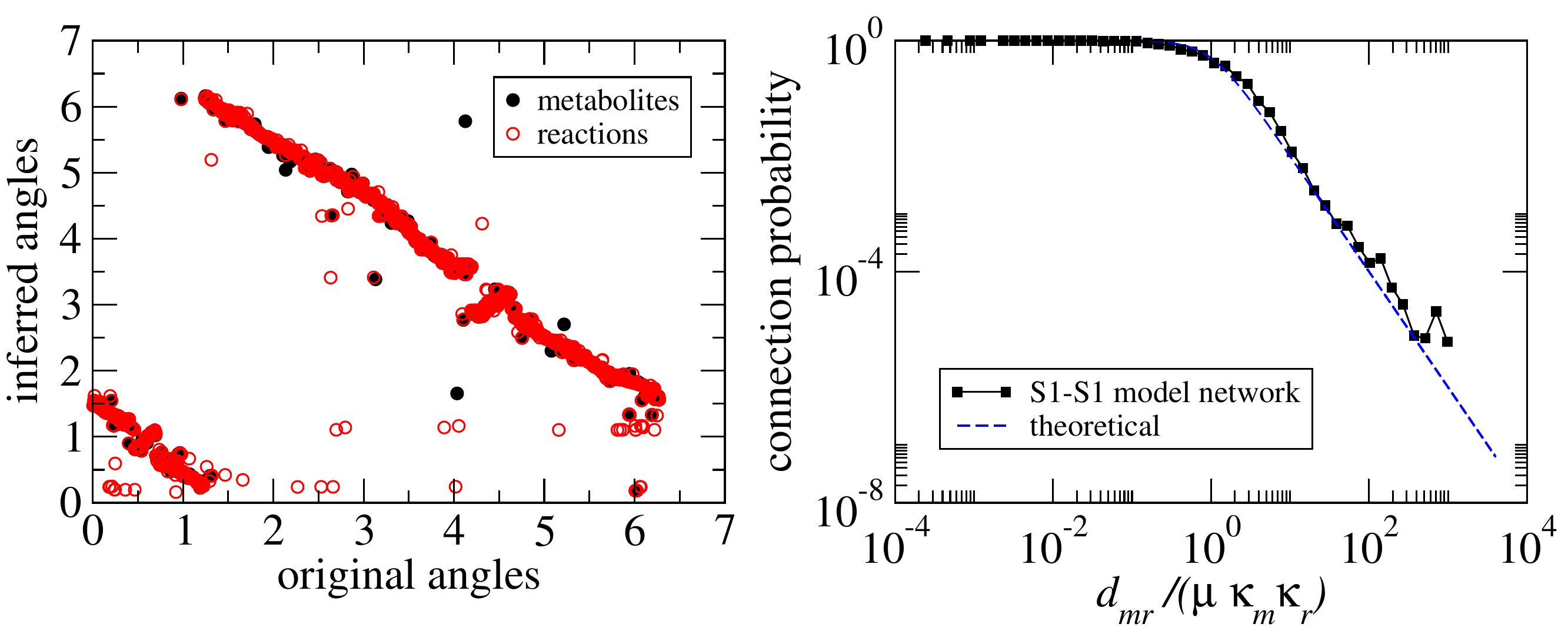}
\caption{{\bf Calibration of the embedding algorithm.} The left plot shows the inferred angular coordinates of metabolites and reactions vs. the real ones of a network generated with the $\mathbb{S}^1 \times \mathbb{S}^1$ with the same parameters as the real metabolism. The right plot shows the empirical connection probability obtained from the embedding compared to the theoretical one in Eq.~(\ref{p(x)}) }
\label{fig:validation}
\end{figure*}

Having found the MLE values for expected degrees $\kappa_m$, we now
have to maximize Eq.~(\ref{likelihood}) with respect to angular
coordinates. This task is equivalent to maximizing the
partial log-likelihood
\begin{eqnarray}
\ln{{\cal L}(\mathbb{A}|\{\kappa_{m,i}^*,\theta_{m,i}, \theta_{r,j}\})}= \nonumber \\
\sum_{i=1}^{N_m} \sum_{j=1}^{N_r} \left\{ a_{ij} \ln{p(x_{ij})}+ (1-a_{ij}) \ln{[1-p(x_{ij})]} \right\}.
\label{loglike}
\end{eqnarray}
The maximization of Eq.~(\ref{loglike}) with respect
to the angular coordinates cannot be performed analytically and we have to rely on numerical optimization procedures. Unfortunately, the low degrees of reactions implies that  any attempt to maximize Eq.~(\ref{loglike}) directly is doomed to fail. Indeed, the uncertainty in the position of a low degree reaction is necessary very high. This, in turn, increases the uncertainty in the position of its metabolites' neighbors, which translates into global uncertainty in the localization of nodes and metabolites. We therefore adopt a different strategy. Starting from the original bipartite network, we construct its one mode projection over the space of metabolites, that is, we consider only one type of nodes (metabolites) and declare two metabolites as connected if they participate in the same reaction in the original bipartite net. If metabolites are power-law distributed in the bipartite network, the obtained unipartite network is also power-law distributed with the same exponent. This solves the problem mentioned above because, now, high degree nodes can be located with high accuracy so that we can use afterwords these nodes as a template to find the coordinates of the rest of the nodes.

We find the angular coordinates of metabolites by fitting the one-mode projected network using the $\mathbb{S}^1$ model as described in~\cite{Boguna:2010fk}. Once the angular coordinates $\theta_{m,i}^*$ are known, we find the optimal angular coordinates of reactions by maximizing Eq.~(\ref{loglike}) but using the already known coordinates of metabolites as fixed inputs. This final maximization is a simple procedure because, being $\theta_{m,i}^*$ fixed, we can maximize the likelihood of each reaction independently.

We first test the described procedure in synthetic networks generated by the $\mathbb{S}^1 \times \mathbb{S}^1$ model with the same parameters as the real E. coli metabolism. Results are shown in Fig.~\ref{fig:validation}. The left plot shows the inferred angles for metabolites and reactions vs. the real ones. As it can be clearly seen, up to minor fluctuations and a global phase shift due to rotational symmetry of the model, the agreement between the real coordinates and those inferred by the algorithm is very good. The right plot shows the connection probability using the inferred coordinates vs. the one used to generate the model Eq.~(\ref{p(x)}). Again, the agreement between the two is excellent.

\section{Classification of pathways in E. coli depending on localization}
See Table I.
\begin{table*}[t]
\caption{Classification of E. coli's pathways. Pathways are classified as ``localized'' (75\% of the pathway localized in a single bin), ``bimodal'' (75\% of the pathway localized in two bins) ``multi-peaked'' (75\% of the pathway localized in three bins or more with at least one peak above 25\%), and ``transversal'' (no bin above 25\%) according to the results and bin size of Fig.~\ref{fig:3}. Pathways in italics indicate that, although they are split in two or three bins, these bins are adjacent and so a change in the bin resolution would lead to their redefinition as more localized pathways.}
\scriptsize
\begin{center}
\begin{tabular}{|c|c|c|c|}
\hline
LOCALIZED              & BIMODAL                & MULTI-PEACKED & TRANSVERSAL\\ \hline
Glu                                    & His                                   &Ala, Asp                                &Cofactor and Prosthetic\\ \hline
Folate                                & Met                                  &Arg, Pro                         &  {\it Purine and Pirimidine}\\ \hline
Methylglyoxal                    & Thr, Lys                            &Cys                 &  Alternate Carbon\\ \hline
Oxidative Phosphorilation  &{\it Anaplerotic}               & Gly, Ser       &Transport Inner Membrane\\ \hline
 Murein B                          &{\it Citric Acid Cycle}        & {\it Tyr, Phe, Trp}                       &\\ \hline
Murein R                          & Glyoxylate                        & Val, Leu, Ile      & \\ \hline
                                         & {\it Pentose Phosphate}  & {\it Nucleotides S}  & \\ \hline
                                         &  Inorganic Ion Transport &  tRNA Charging                      & \\ \hline
                                          & Membrane Lipid            &    Glycolisis                       &  \\ \hline
                                          &                                        &  Pyruvate                           & \\ \hline
                                           &                                       & {\it Nitrogen} & \\ \hline
                                           &                                       & Lipopolysaccharide & \\ \hline
                                           &                                       & {\it Cell Envelope}  & \\ \hline
                                           &                                       & {\it Glycerophospholipid} & \\ \hline
\end{tabular}
\end{center}
\label{table:1}
\end{table*}%

\section{Pathways crosstalk and the disparity filter}
We use the following measure of crosstalk between pathways:
\begin{equation}
XT_{PaPb}=\sum_{j \in P_a}\sum_{j' \in P_b}\sum_{i \in \nu}(p(x_{ij})+p(x_{ij'}))|_{\mbox{observed links}},
\end{equation}
where $\nu \in \mathcal{M}_{ab}$ is the set of metabolites shared by the reactions in the two pathways $P_a$ and $P_b$, and only probabilities of connections associated to observed links are considered.

Of $561$ possible pathway pairs in E. coli, $460$ are non-zero crosstalk ($82.00\%$) with a minimum value of $1.80$ and a maximum of $159.91$. In human cells, of $4278$ possible pathway pairs, $1689$ are non zero ($38.64\%$) with a minimum crosstalk of $1.19$ and a maximum of $131.28$. Moreover, there is an isolated pathway (48, Limonene and Pinene Biosynthesys) without crosstalk (no common metabolites with other pathways). So, at this level human cells metabolism seems to be more modular than E. coli's. 

\begin{figure}[ht]
\includegraphics[width=8.7cm]{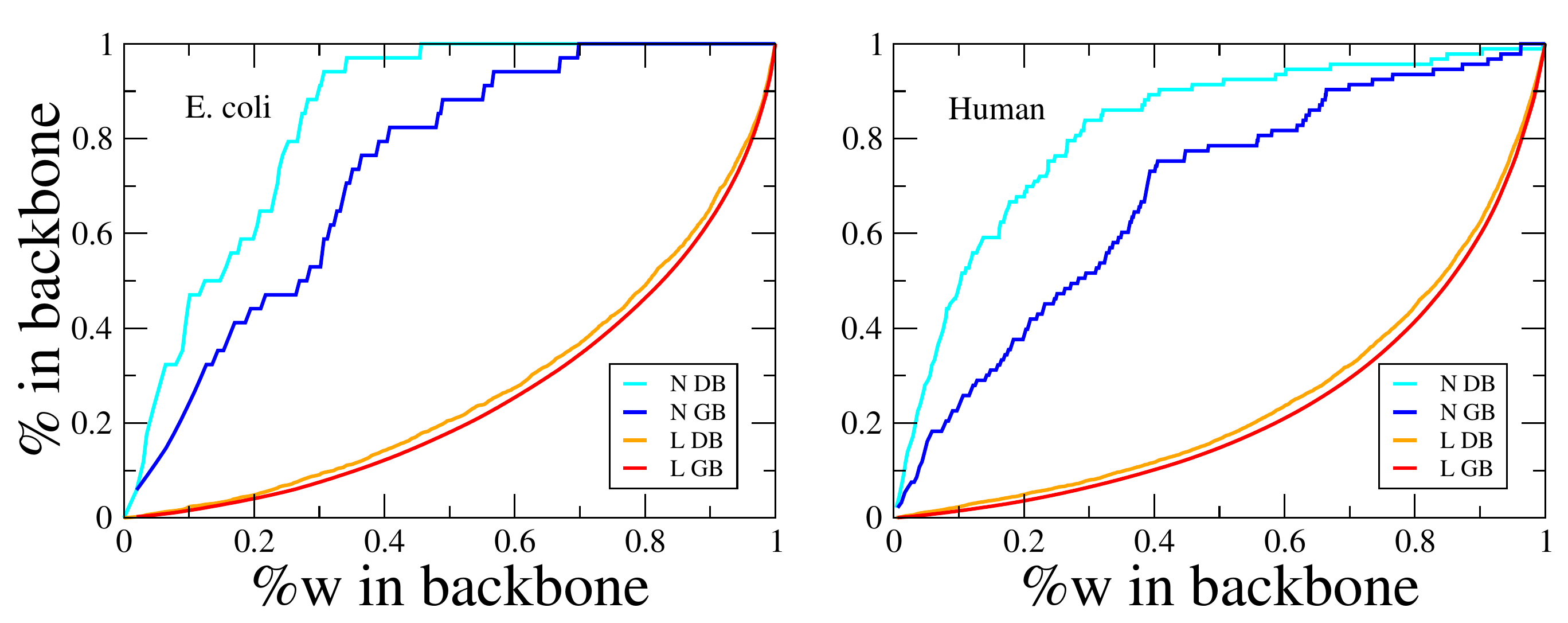}
\caption{{\bf Disparity backbone vs global threshold backbone.}}
\label{fig:pathways2}
\end{figure}
The obtained pathway crosstalk matrices are filtered to obtain backbones according to the multiscale methodology in \cite{Serrano:2009b}, which do not belittle small pathways and gives an effective tradeoff between maximum weight and nodes in the backbone with the minimum number of links. A global threshold filter would lose many more nodes for the same number of links and weight in the backbone, see Fig.~\ref{fig:pathways2}.

The disparity filter methodology preserves interactions with a statistically significant intensity for at least one of the two nodes the edge is incident to. To decide whether a connection is relevant, the filter compares against a null hypothesis which assumes that the local weights associated to a node are uniformly distributed at random. In this way one discounts intensities that could be explained by random fluctuations. More specifically, a $p$ value --the probability $\alpha_{ij}$ that if the null hypothesis is true one obtains a value for the normalized weight $w_{ij}/s_i$ between nodes $i$ and $j$ larger than or equal to the observed one-- is calculated for each edge in the network. By imposing a significance level $\alpha$, the links that carry weights that can be considered not compatible with a random distribution can be filtered out with a certain statistical significance. Links in the backbone will be then those which satisfy
\begin{equation}
\alpha_{ij}=1-(k-1)\int_0^{w_{ij}/s_i} (1-x)^{k-2}dx < \alpha,
\label{eq:confidencelevel}
\end{equation}
where $k$ is the degree of node $i$.
By changing the significance level, we can filter out the links progressively focusing on more relevant ones. As a result, the disparity filter reduces significantly the number of edges in the original network, while keeping almost a large fraction of the total weight and the total number of nodes. It preserves as well the cutoff of the degree distribution, the form of the weight distribution, and the clustering coefficient.

\section{Results for human cells metabolism}
\begin{figure*}[ht]
\includegraphics[width=14cm]{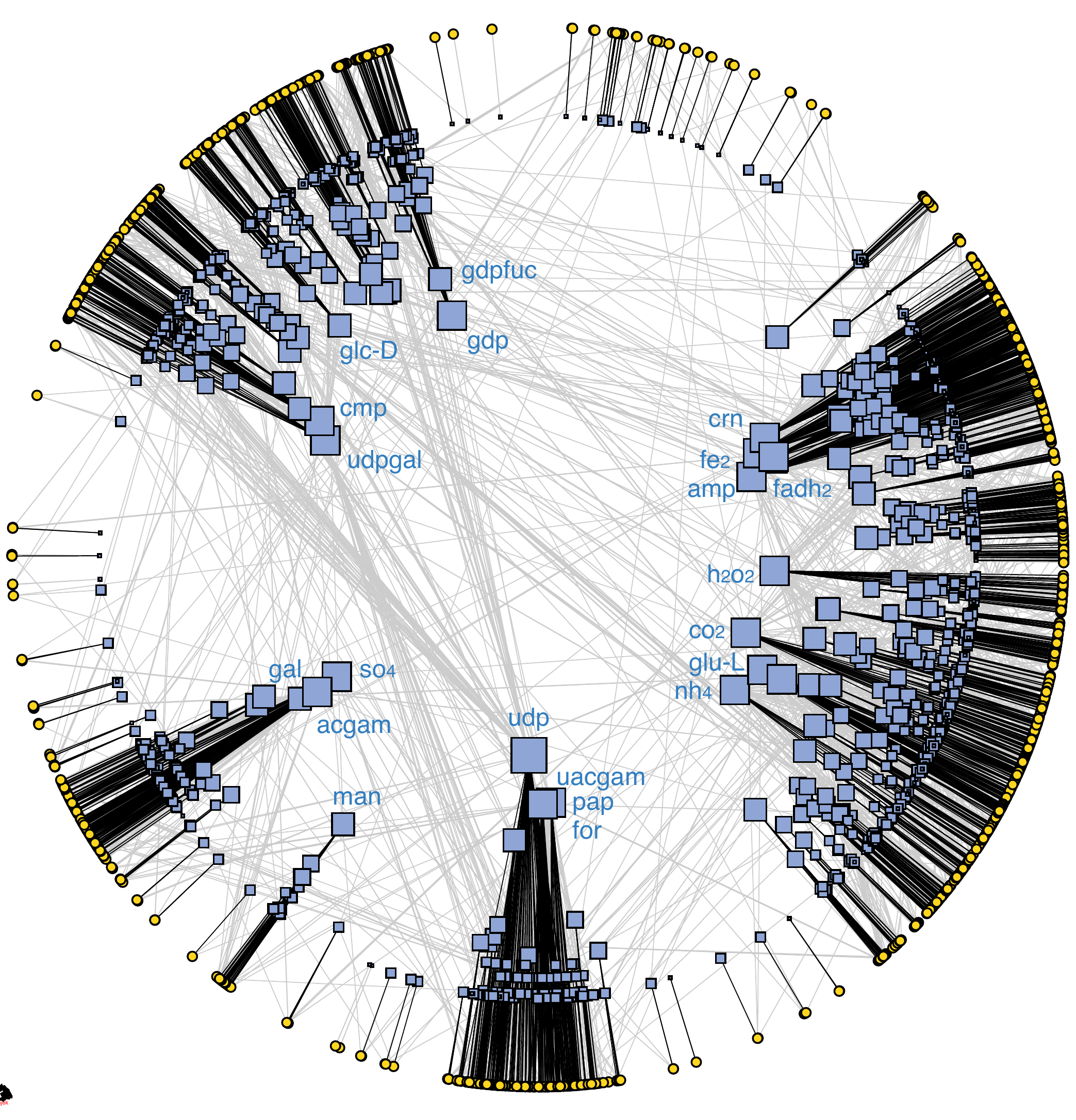}
\caption{{\bf Human metabolism map.} Yellow circles represent reactions whereas blue squares are metabolites. For each metabolite, the symbol size is proportional to the logarithm of the degree and radially placed according to the expression $r = R-2 \ln k_m$. Black (grey) connections are those that according to the model have a probability of existence larger (smaller) than $0.5$.}
\label{fig:humanmap}
\end{figure*}

In Fig.~\ref{fig:humanmap}, we show the embedding representation of human cells metabolism.  In Fig.~\ref{fig:pathways}, we show 
the angular distribution on the ring of the whole list of pathways evaluated from the circle-based embedding of 
the reactions they involve.

\begin{figure*}[ht]
\includegraphics[width=17cm]{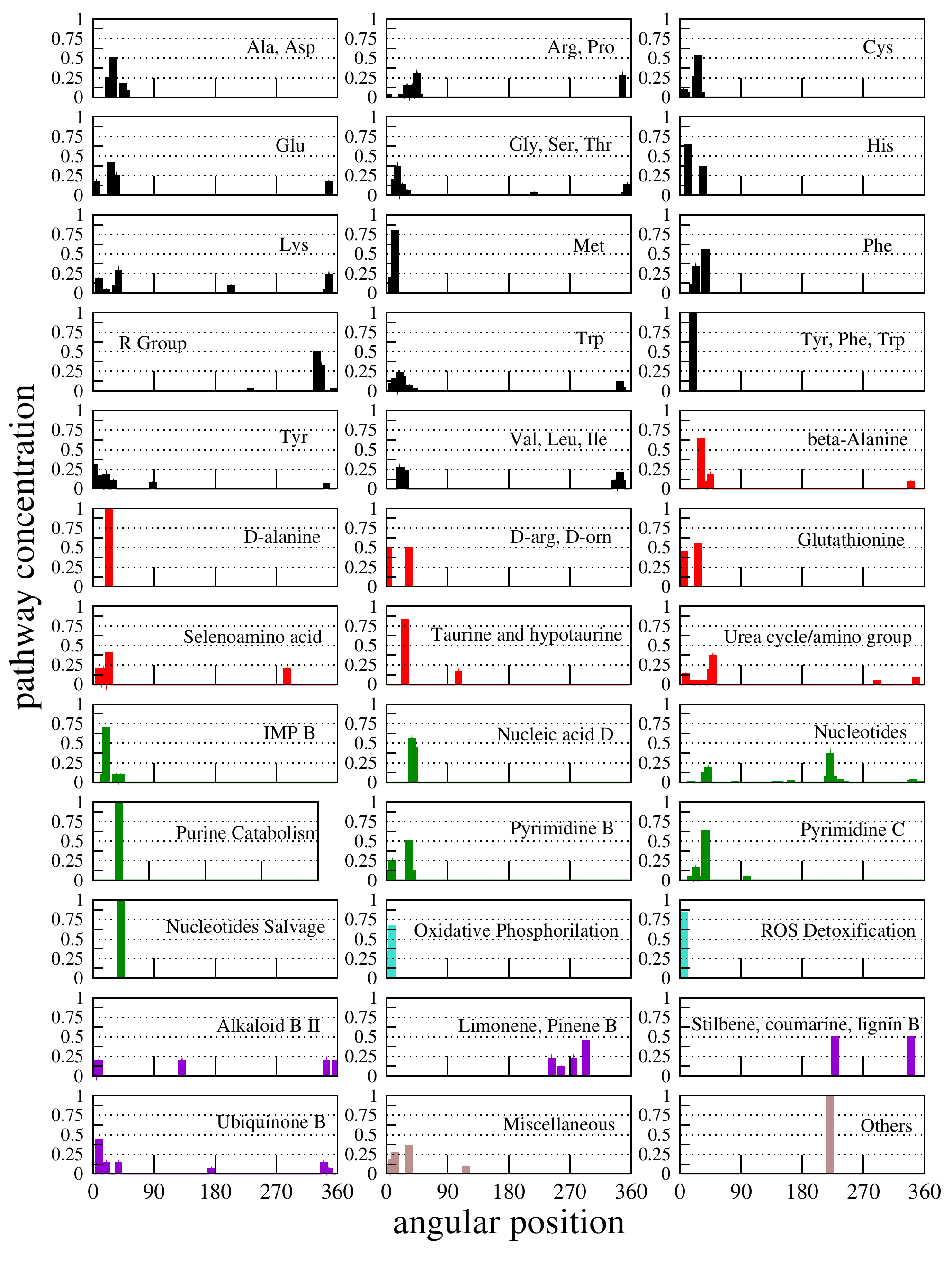}
\end{figure*}
\begin{figure*}[ht]
\includegraphics[width=17cm]{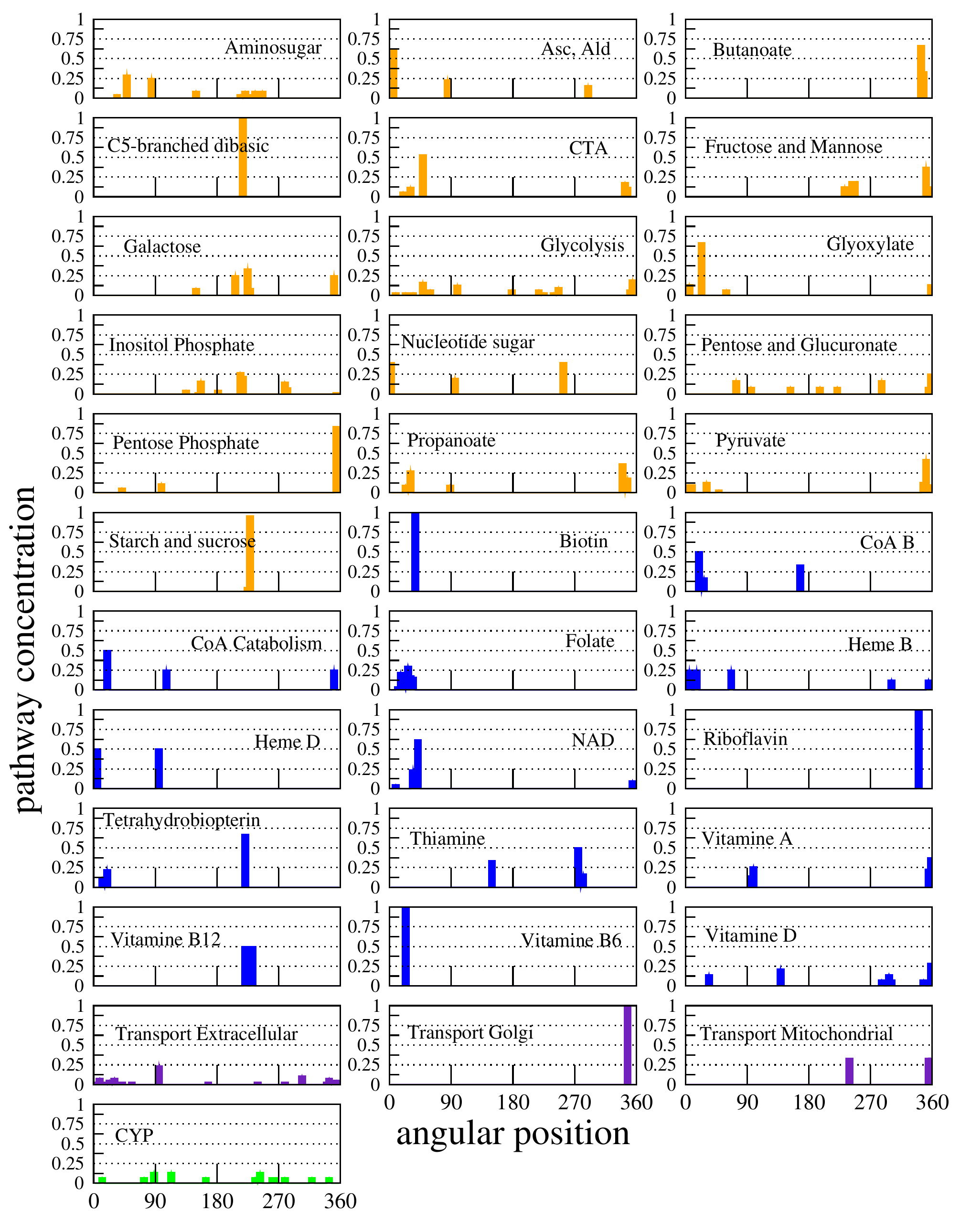}
\end{figure*}
\begin{figure*}[ht]
\includegraphics[width=17cm]{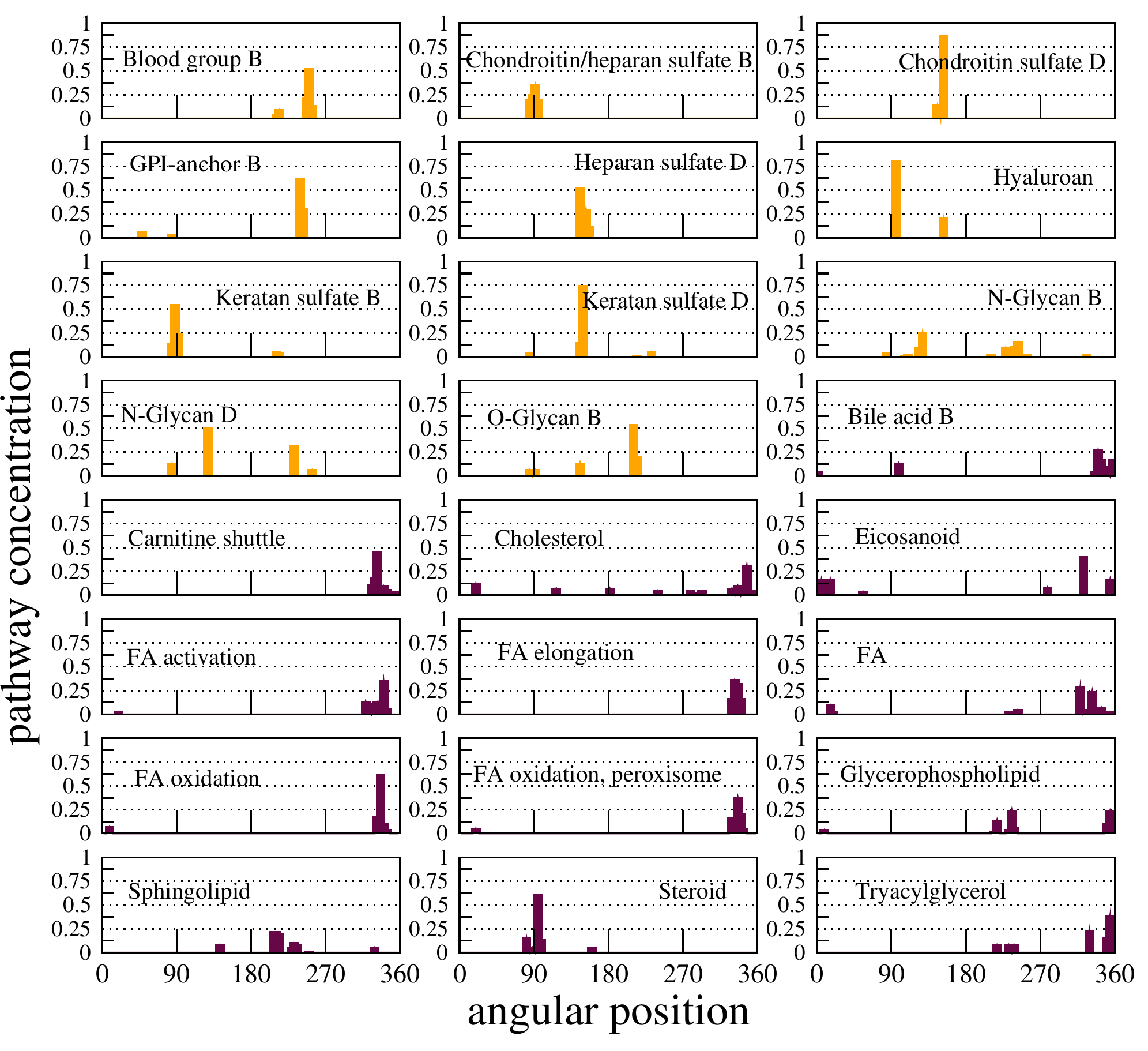}
\caption{{\bf Angular distribution of pathways for the human metabolism.} The whole angular domain $[0,360^o]$ is divided in $50$ bins of $7,2^o$ each and for each bin we compute the fraction of reactions of the pathway in it. Each pathway is shown in a different graph. Different colors indicate different metabolic families. Panel I: black for Amino Acids metabolism (numbering the graphs from left to right and from top to bottom, 1-14), red for metabolism of Other Amino Acids (15-21), dark green for Nucleotide metabolism (22-28), turquoise for Energy metabolism (29,30), purple for biosynthesis of Other Secondary Metabolites (31-34), brown for miscellaneous and others (35,36). Panel II: orange for Carbohydrate metabolism (1-16), blue for metabolism of Cofactors and Vitamins (17-30), violet for Transport pathways (31-33), light green for Xenobiotics Biodegradation (34). Panel III: orange for Glycan metabolism (1-11), and dark brown for Lipid metabolism (12-24). Pathway names have been abbreviated in standard forms whenever possible.}
\label{fig:pathways}
\end{figure*}

\end{document}